\documentclass[aps,prd,twocolumn,10pt,eqsecnum,showpacs,nofootinbib,amsfonts,amssymb,amsmath,floatfix]{revtex4-1}
\usepackage[colorlinks=true,allcolors=blue]{hyperref}
\usepackage{graphicx}
\usepackage{color}

\def\be{\begin{equation}}
\def\ee{\end{equation}}

\begin{document}

\author{\'Eanna \'E.\ Flanagan}
\author{David A.\ Nichols}
\affiliation{Cornell Center for Astrophysics and Planetary Science (CCAPS),
Cornell University, Ithaca, New York 14853 USA}

\title{Observer dependence of angular momentum in general relativity and its relationship to the gravitational-wave memory effect}

\begin{abstract}

We define a procedure by which observers
can measure a type of special-relativistic linear and angular momentum
$(P^a, J^{ab})$ at a point in a curved spacetime using only the spacetime
geometry in a neighborhood of that point.  The method is chosen to yield
the conventional results in stationary spacetimes near future null
infinity.

We also explore the extent to which spatially separated observers
can compare the values of angular momentum that they measure and find consistent results.
We define a generalization of parallel transport along curves which
gives a prescription for transporting values of angular momentum along
curves that yields the correct result in special relativity.
If observers use this prescription,
then they will find that the angular momenta they measure are observer dependent, because of the effects of spacetime curvature.
The observer dependence can be quantified by a kind of generalized holonomy. We show that bursts of gravitational waves with memory generically give rise to a nontrivial generalized holonomy: there is, in this context, a close relation between the observer dependence of angular momentum and the gravitational-wave memory effect.

\end{abstract}

\pacs{04.20.Ha, 04.25.Nx}

\maketitle

\section{Introduction and Summary}

\subsection{Angular momentum in general relativity}

An interesting nonlinear feature of dynamical asymptotically flat solutions
in general relativity is
that there is no canonical way to define a special-relativistic angular
momentum at future null infinity.
This result follows from the work of Bondi, van der Burg, and Metzner
\cite{Bondi1962} and Sachs \cite{Sachs1962,Sachs1962b}.
They showed that the group of asymptotic symmetries of asymptotically
flat spacetimes at future null infinity
is not the Poincar\'e group but is instead an infinite-dimensional 
group now known as the Bondi-Metzner-Sachs (BMS) group.
Its structure is similar to that of the Poincar\'e group:
rather than being a semidirect product of the Lorentz group with a four-parameter Abelian group of spacetime
translations, it is a similar product of the conformal group on a
2-sphere (which is isomorphic to the universal covering group of the Lorentz group) with an
infinite-dimensional commutative group called the supertranslations
\cite{Stewart1993}.
The translations are a four-parameter normal subgroup of the larger group of
supertranslations, from which the Bondi energy-momentum \cite{Bondi1960} is
defined.
The supertranslations, however, make relativistic angular momentum (the charge
associated with the Lorentz symmetries) behave differently in asymptotically
flat spacetimes than in Minkowski space.  In the latter, angular momentum
depends only upon a choice of origin, which is a consequence of the four
spacetime translations in the Poincar\'e group; in the former, angular momentum
depends upon a smooth function on the 2-sphere that parametrizes the
supertranslations in the BMS group.
This property is typically called the supertranslation ambiguity of angular
momentum (e.g., \cite{Ashtekar1979}).  It arises because there is no unique
way to pick out a preferred Poincar\'e group with which to define a
special-relativistic angular momentum.

Instead of special-relativistic linear and angular momentum, one has an infinite set of conserved
charges, one associated with each generator of the BMS group \cite{Dray1984,Wald2000}.
These charges can be computed from a surface integral over any cross section of future null infinity,
and the difference between the values of a charge at two different cross sections is
given by the integral of a 3-form (or ``flux'') over the region of future null infinity between the two
cross sections.  These BMS charges transform covariantly under BMS transformations,
just as special-relativistic linear and angular momentum transform under Poincar\'e transformations,
and they include the Bondi 4-momentum.

Some researchers have argued that it is necessary or desirable to
give a definition of a preferred finite set of conserved charges, that
would be more similar to the familiar conserved charges of special
relativity
\cite{Adamo2012,Moreschi1986,Moreschi1988,1998PhRvL..81.1150R,Dain2000}.
An alternative philosophy, which we espouse, is that all of the BMS
charges are physically relevant and that one should try to understand
more deeply their physical nature, starting with operational
prescriptions by which they can be measured by asymptotic observers.

The purpose of this paper is an initial attempt to understand how BMS
charges can be measured.  For simplicity,
we take a ``bottom-up'' approach: we suppose that observers
who are unaware of the BMS group attempt to measure conserved charges
and ask how the charges they measure are related.  (This is analogous
to Newtonian observers who are ignorant of special relativity making
measurements in Minkowski spacetime; their observations of Newtonian
quantities are inconsistent because of Lorentz contraction, etc.)
Similarly, here we find that the charges measured by different
observers using special-relativistic methods are inconsistent due to
spacetime curvature.  One of our goals is to characterize or interpret the
inconsistencies between different observers' measurements, and
relate the inconsistencies to other measurable quantities.

For simplicity, we will restrict attention to measurements made in
stationary regions, and focus on how measurements made in two
successive stationary regions can be compared to one another.

More specifically, our main results are as follows:
\begin{itemize}

\item We give a local, operational definition
of an ``angular momentum'' that can be measured by an observer at a point in a curved spacetime, using only information contained in the geometry in a neighborhood of that point.  The result is a pair of tensors $(P^a, J^{ab})$ at that spacetime point.
This prescription is chosen to give
the expected result in stationary spacetime regions near future null infinity.  See Sec.\ \ref{sec:AngMom} for details.

\item We define a method by which two observers at two different points in a curved spacetime can compare the values of angular momenta that they measure.  The philosophy we adopt is to imagine observers who assume the validity of special relativity and who make measurements based on this assumption.  We devise a method of comparison based on a generalization of parallel transport, which reduces to the correct method in flat spacetimes.  In curved spacetimes, the method of comparison will be curve dependent, and,  in general, inconsistencies will arise when observers attempt to compare values of angular momenta.  Therefore, from this point of view, angular momentum inevitably becomes observer dependent in curved spacetimes.  See Sec.\ \ref{sec:GenHolDef} for details.

\item We identify a simple physical mechanism that accounts for and explains the observer dependence in simple cases.
Specifically, two observers who measure the change in angular momentum
of a given source can disagree on that change, since they disagree on
where they believe the source used to be.  They disagree on the
source's original location because of the gravitational-wave memory
effect, the permanent relative displacement of observers due to the
passage of a burst of waves \cite{Zeldovich1974,
Christodoulou1991,BieriGarfinkle2014}.  The memory effect has,
unbeknownst to the observers, displaced them by different amounts.  This argument is
given in more detail in Sec.\ \ref{subsec:NewtArg} below.

\item We argue that the close relation between gravitational-wave
memory and the observer dependence of angular momentum is in fact
very general, by using covariant methods and looking at a number of
examples (Secs.\ \ref{sec:GenHolDef} and \ref{sec:GenHolLin}).
While the connection between gravitional-wave memory and angular-momentum
ambiguity has often been noted, our analysis shows the explicit and
precise form of the relationship in a general context.
In addition, there is a close relation between gravitational-wave memory and the BMS supertranslation that relates the shear-free cuts of a stationary spacetime before a burst of gravitational waves to those after the burst, as noted by, for example, Strominger and Zhiboedov \cite{Strominger2014}.  
Therefore, our examples in Secs.\ III and IV also highlight the role of BMS supertranslations in the observer dependence of angular momentum in these simple contexts.

\end{itemize}

\subsection{Universality of observer dependence of angular momentum}
\label{subsec:NewtArg}

As discussed above, if different observers attempt to measure angular momentum
in general relativity using special-relativistic methods, they will disagree on the results.
In other words, angular momentum becomes observer dependent.  In this section,
we will show that this observer dependence is a universal
and local feature of general relativity, independent of the choice of
asymptotic boundary conditions.  (However, this observer dependence
does not necessarily imply the existence of ambiguities of the BMS
type, as we discuss in Sec.\ \ref{sec:Conclusions}).
We will compute the observer dependence explicitly in a specific
simple case, and show that it is closely related to gravitational-wave memory.

Consider two observers $A$ and $B$ (Alice and Bob) in a flat region of spacetime,
who are at rest with respect to one another and share a common inertial frame $(t, \bf{x})$.  Suppose that they both measure the angular momentum of a nearby particle.  The observer $A$ will obtain the result
\be
{\bf J}_A = {\bf S} + ({\bf x}_p - {\bf x}_A)\times {\bf p},
\ee
where ${\bf S}$ is the intrinsic angular momentum of the particle, ${\bf p}$ 
is its momentum in the observers' common inertial frame, ${\bf x}_p$ is the 
location of the particle, and ${\bf x}_A$ is $A$'s location.
Here we assume that $A$ measures the angular momentum about her own location.  Similarly, observer $B$ will measure an angular momentum about his own location, and obtain the result
\be
{\bf J}_B = {\bf S} + ({\bf x}_p - {\bf x}_B)\times {\bf p}.
\ee
If $A$ and $B$ compare their measurements, they will find a difference given by
\be
{\bf J}_A - {\bf J}_B = - ({\bf x}_A - {\bf x}_B) \times {\bf p},
\ee
which is consistent with their measured relative displacement ${\bf x}_A - {\bf x}_B$.

Now suppose that a gravitational-wave burst of finite duration is incident on the observers.  Thus the spacetime consists of a flat region, followed by a gravitational-wave pulse, followed by a subsequent flat region.  We can adopt transverse-traceless (TT) coordinates $(T,X^i)$ to describe the entire relevant spacetime region---before, during, and after the burst of waves---which we chose to coincide with the inertial-frame coordinates $(t,x^i)$ before the burst.  In the TT coordinates and in the linearized approximation, the metric is
\be
ds^2 = - dT^2 + [\delta_{ij} + h_{ij}(T - Z) ] dX^i dX^j,
\label{metric00}
\ee
where for simplicity we have specialized to a burst propagating in the $+Z$ direction.  At late times, the metric perturbation becomes constant,
$h_{ij}(T - Z) \to h^\infty_{ij} = $ (constant), while at early times $h_{ij}$ vanishes, and $T=t$, $X^i = x^i$.

We now extend the definition of the coordinates $(t,x^i)$ to the region after the burst by defining
\be
t = T, \ \ \ \
x^i = \left[\delta_{ij} + \frac{1}{2} h^\infty _{ij} \right] X^j.
\label{inertialcoords}
\ee
These coordinates are then inertial coordinates after the burst.
Now the observers $A$ and $B$ are freely falling, which implies that their TT
coordinate locations $X_A^i$ and $X_B^i$ are conserved.  Hence their
relative displacement in the $(t,x^i)$ inertial frame after the burst
is
\be
{\bf{x}}^\prime_A - {\bf{x}}^\prime_B  = \left[ {\bf 1} + \frac{1}{2} {\bf h}^\infty \right]\cdot
({\bf{x}}_A - {\bf{x}}_B).
\label{mem}
\ee
This is the standard formula for gravitational-wave memory.  Here
$\bf{x}^\prime_A$ and $\bf{x}^\prime_B$
are the locations of $A$ and $B$ in the inertial-frame coordinates after the burst.

In the spacetime region after the burst has passed, the observers A and B can again measure the angular momentum of the particle, in the new inertial frame
$(t,x^i)$. The observer $A$ obtains the result
\be
{\bf J}^\prime_A = {\bf S}^\prime + ({\bf x}^\prime_p - {\bf x}^\prime_A)\times {\bf p}
\, ,
\ee
where primes denote quantities as measured after the burst.  We imagine that the particle's spin and location may have changed in the intervening period, but for simplicity we assume that its momentum ${\bf p}$ has not.  A similar formula applies to the observer $B$, and once again if
$A$ and $B$ compare their measurements, they will find a difference
given by
\be
{\bf J}^\prime_A - {\bf J}^\prime_B = - ({\bf x}^\prime_A - {\bf x}^\prime_B) \times {\bf p},
\ee
which is consistent with their measured relative displacement
after the burst ${\bf x}^\prime_A - {\bf x}^\prime_B$.  So far, there
is no observer dependence.

Next, we assume that observer $A$ wishes to compute the change in
angular momentum of the particle between early and late times.
This is given by
\be
\delta {\bf J}_A = {\bf J}_A^\prime - {\bf J}_A + \delta {\bf x}_A
\times {\bf p}.
\label{11}
\ee
Here $\delta {\bf x}_A$ is the change in $A$'s location between early
and late times, and the third term is necessary to transform the
original angular-momentum measurement to her new location, so that she
is subtracting angular momenta as measured about the same point.
However, as far as observer $A$ is concerned, $\delta {\bf x}_A$
vanishes, since she is an inertial observer sitting at the origin of
her inertial frame.  In particular, she is unaware of the effects of
the gravitational-wave burst.  (More generally, if the observer were
accelerated by nongravitational forces, she could measure
$\delta {\bf x}_A$ using an accelerometer carried with her.  In the
present context, the accelerometer reading would be zero.)

Inserting the assumption $\delta {\bf x}_A = 0$ into Eq.\ (\ref{11})
and subtracting a similar equation for $B$ finally yields
\be
\delta {\bf J}_B - \delta {\bf J}_A = ({\bf x}_B - {\bf x}_A)\times {\bf p} -
({\bf x}'_B - {\bf x}'_A) \times {\bf p} \;.
\ee
Using the gravitational-wave-memory formula (\ref{mem}) simplifies this
to
\begin{equation}
\label{eq:NewtJdiff}
\delta {\bf J}_B - \delta {\bf J}_A = - \frac 12 [\boldsymbol{\mathsf h}^\infty
\cdot ({\bf x}_B - {\bf x}_A)] \times {\bf p} \; .
\end{equation}
Thus, $A$ and $B$ disagree on the change in angular momentum, by an
amount which is proportional to the gravitational-wave memory.
Essentially what has happened is that the two observers disagree on
where the particle used to be, because they have been displaced
relative to one another by the gravitational-wave memory effect, and
they assume there is no such relative displacement.

The result (\ref{eq:NewtJdiff}) will be rederived by a more formal and
covariant computation in Sec. \ref{subsec:GenHolNewt} below.

\subsection{Covariant description of angular momentum's observer dependence:
Methods of this paper}

While the example of the previous section was intuitively
useful and suggestive of the generality of the phenomenon, it is important
to have a covariant method for comparing angular momenta
at different times
and as measured by different observers.
There are, however, several subtle aspects of how to define angular momentum
and how to compare values between different observers in curved spacetime.
The remainder of this paper is devoted to articulating a procedure that
treats these issues and allows observers to compare angular momentum
covariantly.
We now give a brief sketch of the approach taken in this paper and
summarize the organization of
this paper's sections.

Section \ref{sec:AngMom} contains a local operational definition of a
linear and angular momentum $(P^a,J^{ab})$ that can be measured by
individual observers.
For simplicity we will refer to this pair simply as ``angular momentum''.
Section \ref{subsec:AngMomDef}
defines the mathematical space in which the local angular momentum lives:
the dual space of the space of Poincar\'e transformations from the
tangent space at a point in spacetime to itself.
The next part, Sec.\ \ref{subsec:AngMomStat}, defines a prescription
whereby an angular momentum (a particular element of this dual
space) can be obtained from local measurements of the Riemann tensor and its derivatives.
Section \ref{subsec:AngMomDefMotivation} shows that the prescription for
measuring angular momentum yields the expected value in stationary
spacetimes near future null infinity, Sec.\ \ref{subsec:interp} describes the
accuracy and errors of the algorithm in more general spacetimes, Sec.\
\ref{subsec:nonunique} notes that the algorithm is not unique, and Sec.\
\ref{subsec:AngMomCOM} focuses on the accuracy with which the center-of-mass
worldline can be measured.

Section \ref{sec:GenHolDef} describes how to compare angular momentum at
different spacetime points.
It defines a transport law in Sec.\ \ref{subsec:AffineTrans}---which will be
called the {\it affine transport}---that can be used for comparing angular
momentum at two different spacetime points.
Section \ref{subsec:GenHolApps} explains in detail how to compare angular
momentum at two points using the affine transport.
When the curve is a closed loop, the transport law defines a {\it generalized
holonomy} operation, the basic properties of which are given in Sec.\
\ref{subsec:GenHol}.
When the generalized holonomy reduces to the identity, it indicates that there
is a consistent (observer-independent) notion of angular momentum for
different observers along the curve; when it does not, it provides a notion
of the size of the observer dependence in angular momentum between different
observers along the curve.
Section \ref{subsec:GenHolFermi} shows that the generalized holonomy
contains four independent pieces.  When the closed curve is generated
by portions of worldlines of two freely falling observers, connected
by spatial geodesics, each of the four pieces can be interpreted as
a kind of gravitational-wave memory.  In particular, for nearby
geodesics, the generalized holonomy contains the usual gravitational-wave 
memory.

Section \ref{sec:GenHolLin} gives two examples of the generalized holonomy for
an idealized spacetime consisting of a region of flat Minkowski space
followed by a burst of linearized gravitational waves with memory that
propagates away leaving a second flat Minkowski spacetime region.
The first half of the section, Sec.\ \ref{subsec:GenHolNewt}, reproduces the
nearly Newtonian argument of Sec.\ \ref{subsec:NewtArg} using the language of
the generalized holonomy.
The next half of the section, Sec.\ \ref{subsec:GenHolGen}, examines the more
general example of a gravitational wave expanded in symmetric trace-free
multipoles that is emitted radially outward from a pointlike source.
The paper concludes in Sec.\ \ref{sec:Conclusions}.

Throughout this paper, we use units in which $G=c=1$, and we use the
conventions of Misner, Thorne, and Wheeler \cite{Misner1973} for the metric
and curvature tensors.
We use Latin letters from the beginning of the alphabet for general
spacetime indices and Greek letters for those associated with specific
coordinate systems.
Latin letters from the middle of the alphabet (starting at $i$) will be
reserved for spatial indices, and a $0$ will denote a time index in the
latter context.

\section{Operational definition of the angular momentum of a source as measured by a local observer}
\label{sec:AngMom}

In this section, we describe a method by which an observer in the vicinity of
some source of gravity can attempt to measure the angular momentum of that
source, by using only information about the geometry of spacetime in the
observer's vicinity.
Specifically, we describe an algorithm by which an angular momentum can
be constructed from the Riemann tensor and its gradients at the
observer's location.
The algorithm we propose, moreover, is not unique, and the angular
momenta obtained will differ from one observer to another.  However, in
a certain limit (Sec.\ \ref{subsec:AngMomDefMotivation} below), the
angular momenta will become observer independent and characterize the
source.
In more general situations,
the nonuniqueness of the algorithm will be unimportant, and the
angular momenta will be observer dependent.  In these situations, the nature of this observer dependence will
be physically interesting, as discussed in the remaining sections of
this paper.

While the literature on angular momentum in general relativity
is extensive and well developed (see, e.g., \cite{Szabados2009}),
our approach here introduces a new perspective, in that it focuses
on a local and operational definition of a quantity that observers 
can measure.  
Our procedure can be applied at any point in any spacetime (subject to a 
small number of local assumptions) and yields the expected result in the 
limit of large distances from a
source in an isolated, linearized, stationary, vacuum spacetime.

A measurement of the general type considered here, where the angular
momentum of a source is extracted from measurements of the geometry of
spacetime, has been carried out once in the history of physics: the
measurement of the spin of the Earth to $\sim 20$\% by Gravity Probe
B \cite{PhysRevLett.106.221101}.\footnote{Our measurement procedure is not quite the same as
that used by Gravity Probe B.  While both extract angular-momentum
information from the spacetime geometry,
our procedure uses the curvature tensors in an infinitesimal region about a
spacetime point, while Gravity Probe B uses information about the geometry in the
vicinity of an entire orbit in addition to information
about asymptotic inertial frames provided by the direction to a guide star.}

We start in Sec.\ \ref{subsec:AngMomDef} by defining a vector space that can
be interpreted as the space of angular momenta for an observer at a
given point $\mathcal P$ in a curved spacetime.
We give the general algorithm for measuring angular momentum in Sec.\ \ref{subsec:AngMomStat}.
In Sec.\ \ref{subsec:AngMomDefMotivation} we explain the motivation for this
algorithm: namely, that it gives the expected result in stationary linearized
spacetimes near future null infinity.
We discuss the physical interpretation of the measured angular momentum in
general spacetimes in Sec.\ \ref{subsec:interp}, the nonuniqueness of the
algorithm in Sec.\ \ref{subsec:nonunique}, and the accuracy of the
center-of-mass measurement in Sec.\ \ref{subsec:AngMomCOM}.

\subsection{Definition of a linear space of angular momenta at a given
  point in spacetime}
\label{subsec:AngMomDef}

At a point $\mathcal P$ in a spacetime $(M,g_{ab})$, let
$T_{\mathcal P}(M)$ denote the tangent space.
Let $G_{\mathcal P}$ be the Poincar\'e group that acts on
$T_{\mathcal{P}}(M)$, that is, the space of affine maps from
$T_{\mathcal P}(M)$ to itself that preserve the metric.
Since $G_{\mathcal P}$ is a Lie group, it has an associated Lie
algebra $\mathcal G_{\mathcal P}$ that consists of infinitesimal
Poincar\'e transformations.
The corresponding dual space $\mathcal G_{\mathcal P}^*$, the space of linear
maps from $\mathcal G_{\mathcal P}$ to the real numbers, is the space
of linear and angular momenta at the event $\mathcal P$.

To see this explicitly, consider an affine coordinate system
$x^a$ on $T_{\mathcal P}(M)$.
Such a coordinate system is associated with a choice of basis
vectors ${\vec e}_a$ and a fixed vector ${\vec x}_0$ such that
the coordinates $x^a$ of a vector ${\vec x}$
are given by
${\vec x} = {\vec x}_0 + x^a {\vec e}_a$.
In this coordinate system, the maps in $G_{\mathcal P}$ have the usual form
of a Poincar\'e transformation: $x^a \rightarrow \Lambda^a{}_b x^b + \kappa^a$.
Here $\Lambda^a{}_b$ is a Lorentz transformation and $\kappa^a$ is a translation.
The infinitesimal versions of these maps in
$\mathcal G_{\mathcal P}$ have the same form, but with infinitesimal
$\kappa^a$ and with $\Lambda^a{}_b = \delta^a{}_b + \omega^a{}_b$, where $\omega_{ab}$ is an
infinitesimal antisymmetric tensor.
Now consider the dual space, $\mathcal G_{\mathcal P}^*$.
A general linear map from $\mathcal G_{\mathcal P}$ to real numbers can be
written as
\begin{equation}
(\kappa_a,\omega_{ab}) \rightarrow  P^a \kappa_a - \frac{1}{2} J^{ab} \omega_{ab}
 \;
\label{eq:eldef}
\end{equation}
for some vector $P^a$ and some antisymmetric tensor $J^{ab}$.  Therefore, elements of
$\mathcal G_{\mathcal P}^*$ can be parametrized in terms of pairs of
tensors $(P^a,J^{ab})$, a linear momentum and an angular momentum.
The angular momentum $J^{ab}$ transforms under changes of origin in
$T_{\mathcal P}(M)$ as angular momentum should: for
${\vec x}_0 \rightarrow {\vec x}_0 + {\vec {\delta x}}$,
$J^{ab} \rightarrow J^{ab} + 2 P^{[a}\delta x^{b]}$.
The angular momentum $J^{ab}$ would be interpreted by an observer at
${\cal P}$ as angular momentum about a point which is
``displaced from
${\cal P}$ by an amount ${\vec x}_0$,'' even though such a
displacement is ambiguous in general relativity.

\subsection{Definition of the general prescription for measuring an angular momentum}
\label{subsec:AngMomStat}

In this section, we define a prescription for how an observer at an
event $\mathcal P$ can measure an element of the dual space
${\mathcal G}_{\mathcal P}^*$ of linearized Poincar\'e transformations
on the tangent space at $\mathcal P$.
The prescription requires several assumptions about the geometry near
$\mathcal P$, as discussed further below, and therefore it is applicable only in
certain situations.

The steps of the prescription are as follows:

\begin{enumerate}

\item Measure all the components of the Riemann tensor
  $R_{abcd}$ and of its gradient $\nabla_a R_{bcde}$ at
  the event
  $\mathcal P$.  The electric pieces of the Riemann tensor in the observer's
  frame can be measured by monitoring the relative acceleration of
  test masses using the geodesic deviation equation.  Similarly, the
  magnetic pieces can be measured by monitoring the relative angular velocity
  of gyroscopes induced by frame dragging \cite{Nichols2011}.
  By repeating these measurements at nearby spacetime points, the observer can
  in principle also measure the components of the gradient $\nabla_a R_{bcde}$.

\item Compute the curvature invariants
\begin{subequations}
\begin{eqnarray}
K_1 &\equiv& R_{abcd} R^{abcd} , \\
\mathcal K_1 &\equiv & \nabla_a R_{bcde} \nabla^a R^{bcde} .
\end{eqnarray}
\label{eq:invariants}
\end{subequations}
We assume that $K_1 >0$ and $\mathcal K_1>0$.
Then, compute quantities $M$ and $r$ using
\begin{subequations}
\begin{eqnarray}
M &=& \frac{15 \sqrt{5} K_1^2}{4\mathcal K_1^{3/2}}
\;,\\
r &=& \sqrt{\frac{15 K_1}{\mathcal K_1}}
\; .
\end{eqnarray}
\label{eq:rdef}
\end{subequations}

\item Repeat the above measurements and computations at
  nearby\footnote{Equivalently, measure the Riemann
    tensor and its first two derivatives at ${\mathcal P}$; the quantity
  $\nabla_a r$ can then be expressed in terms of these using Eqs.\
 (\ref{eq:invariants}) and (\ref{eq:rdef}).}
  spacetime points, thus measuring the gradient $\nabla_a r$ of the
  quantity $r$.

\item Assuming that the vector $\nabla_a r$ is spacelike,
define the unit vector $n^a$ in the direction of $\nabla_a r$ by
$n^a = N^{-1} \nabla^a r$ where $N = \sqrt{\nabla^a r \nabla_a r}$.
Compute the quantity
\be
y^a = - r n^a
\label{eq:ydef}
\ee
which the observer interprets as the displacement vector from her own
location to the center-of-mass worldline of the source.

\item Compute the symmetric tensor $H_{ab}$ from
\be
H_{ab} = R_{acbd} n^c n^d.
\label{eq:Hdef}
\ee
Compute the eigenvectors $\zeta^a$ and eigenvalues $\lambda$ of this
matrix from $H_{ab} \zeta^b = \lambda \zeta_a$.  From the definition
(\ref{eq:Hdef}), one of the eigendirections will be $\zeta^a = n^a$ with
corresponding eigenvalue $\lambda=0$.  We assume that there is at
least one eigenvector with a strictly positive eigenvalue, and we
denote the eigendirection
corresponding to the largest eigenvalue by $t^a$.  It follows that
this vector is orthogonal to $n_a$, $t^a n_a =0$.

\item Assuming that the vector $t^a$ is timelike, define a
  unit, future-directed timelike vector $u^a$ by $u^a = N^{-1} t^a$
  where $N^2 = -t_a t^a$ and the sign of $N$ is chosen so that $u^a$
  is future directed.  The linear momentum is then given by
  $P^a = M u^a$.

\item Compute the curvature invariant
\begin{equation}
K_2 \equiv \frac{1}{2} \epsilon_{abcd} R^{ab}_{\ \ \ ef} R^{cdef}.
\label{eq:K2def}
\end{equation}
From this compute a spin vector $S^a$ by
\be
S^a = \frac{r^7 K_2}{288 M^2} n^a + \frac{1}{3} r^4 \epsilon^{abcd}
u_b n_c H_{de} u^e.
\label{eq:Sadef}
\ee

\item Compute the angular momentum $J^{ab}$ by
\be
J^{ab} = \epsilon^{abcd} u_c S_d + y^a
P^b - y^b P^a.
\label{eq:Jabdef}
\ee
Finally from $(P^a, J^{ab})$ compute an element of $\mathcal
G_{\mathcal P}^*$ using the definition (\ref{eq:eldef}) specialized to ${\vec
  x}_0=0$.

\end{enumerate}

Although the procedure is somewhat lengthy, these eight steps define a
method for computing an element of $\mathcal G_{\mathcal P}^*$ from the Riemann tensor and
its derivatives at a point $\mathcal P$.

\subsection{Motivation for the prescription: Stationary linearized
  spacetimes near future null infinity}
\label{subsec:AngMomDefMotivation}

We now explain the motivation for the choice of prescription described
in the last subsection: it is designed to give the expected answer
in a certain limit.
Specifically, we consider spacetimes that are stationary and free of matter in the
neighborhood of an observer and for which the sources are sufficiently
distant from the observer that the metric can be described by a
linearized multipolar expansion.  For these distant sources, the
dominant terms in the multipolar expansion will be the mass monopole
and the current dipole or spin, with the remaining multipoles being negligible.
In this situation, the measured $P^a$ and $J^{ab}$
coincide with the conserved charges of the spacetime to a good
approximation, as we now show.  This requirement does not fix the prescription
uniquely, but we shall argue in Sec.\ \ref{subsec:nonunique} below
that the nonuniqueness is not significant.

We start by writing down a Poincar\'e covariant expression for the
metric for stationary linearized spacetimes, keeping only the first
two multipoles.  This metric can be written as
$ds^2 =  (\eta_{\alpha\beta} + h_{\alpha\beta}) dx^\alpha dx^\beta$,
where we have specialized to Lorentzian coordinates $x^\alpha$ for the
background metric, and indices are raised and lowered with $\eta_{\alpha\beta}$.
Let the 4-momentum of the source be ${\hat P}^\alpha =
{\hat M} {\hat u}^\alpha$, where ${\hat u}^a$ is the 4-velocity and
${\hat M}$ is the rest mass.
(We use a hatted notation for these quantities to distinguish them from the
quantities, defined in the previous subsection, that the observer measures.)
Let the intrinsic angular momentum of the
source be ${\hat S}^{\alpha}$ with ${\hat S}^\alpha {\hat u}_\alpha=0$, and
let ${\hat z}^\alpha$ be a point on the center-of-mass worldline of the source.
Let $x^\alpha$ be the point at which we want to evaluate the metric perturbation $h_{\alpha\beta}$.
We define the projection tensor
\be
{\hat p}_{\alpha\beta} = \eta_{\alpha\beta} + {\hat u}_\alpha {\hat u}_\beta
\ee
and define the distance ${\hat r}$ by
$
{\hat r}^2 = {\hat p}_{\alpha\beta} (x^\alpha - {\hat z}^\alpha) (x^\beta - {\hat z}^\beta).
$
Finally, we define the unit vector ${\hat n}^\alpha$ by
\be
{\hat n}_\alpha = \nabla_\alpha {\hat r} = \frac{1}{{\hat r}} {\hat p}_{\alpha\beta} (x^\beta - {\hat z}^\beta).
\ee

In terms of these quantities, the total angular momentum
${\hat J}^{\alpha\beta}$ about the point $x^\alpha$ is
\be
{\hat J}^{\alpha\beta} = \epsilon^{\alpha\beta\gamma\delta} {\hat u}_\gamma {\hat S}_\delta + {\hat y}^\alpha
{\hat P}^\beta - {\hat y}^\beta {\hat P}^\alpha,
\label{eq:hatJabdef}
\ee
where ${\hat y}^\alpha = {\hat p}^\alpha_{\ \beta} ({\hat z}^\beta - x^\beta) = - {\hat r} {\hat n}^\alpha$ is a vector
which points from the field point $x^\alpha$ to the center-of-mass worldline.
The metric perturbation is
\be
h_{\alpha\beta}(x^\alpha) =  \frac{2{\hat M}}{{\hat r}}(\eta_{\alpha\beta} + 2 {\hat u}_\alpha {\hat u}_\beta) - \frac{4}{{\hat r}^2}
{\hat u}_{(\alpha} \epsilon_{\beta)\gamma\delta\epsilon} {\hat S}^\gamma {\hat n}^\delta {\hat u}^\epsilon \; ,
\label{eq:hlinans}
\ee
which is equivalent to the stationary limit of the
linearized metric perturbation in \cite{Thorne1980}, after one makes
the substitution that $u_\alpha = (dt)_\alpha + v_i {\delta^i}_\alpha$,
where $(t,x^i)$ are harmonic coordinates.
Finally, the Riemann tensor is
\be
R_{\alpha\beta\gamma\delta} =  \frac{1}{2}(h_{\alpha\delta,\beta\gamma} + h_{\beta\gamma,\alpha\delta} - h_{\alpha\gamma,\beta\delta} - h_{\beta\delta,\alpha\gamma})
\; ,
\label{eq:linR}
\ee
where
\begin{align}
\label{eq:linh}
& h_{\alpha\delta,\beta\gamma} =  \frac{2{\hat M}}{{\hat r}^3}(\eta_{\alpha\delta}+2{\hat u}_\alpha {\hat u}_\delta)(3{\hat n}_\beta {\hat n}_\gamma-{\hat p}_{\beta\gamma})
  \\
& -\frac{12}{{\hat r}^4} {\hat u}_{(\alpha}\epsilon_{\delta)\lambda\mu\nu} {\hat S}^\lambda {\hat u}^\nu[ (5 {\hat n}_\beta {\hat n}_\gamma -
{\hat p}_{\beta\gamma}){\hat n}^\mu - 2 {\hat n}_{(\beta} {\hat
  p}_{\gamma)}{}^\mu ] \; . \nonumber
\end{align}

We now compute the angular momentum that an observer at $x^\alpha$
would measure in this spacetime, using the algorithm described in the
last subsection.  The curvature invariants
(\ref{eq:invariants})
are given by

\begin{subequations}
\begin{align}
K_1  = & \frac{48{\hat M}^2}{{\hat r}^6} \left[ 1 + O(\epsilon)
 \right]\; , \\
\mathcal K_1 = &
\frac{720{\hat M}^2}{{\hat r}^8}
\left[1+O(\epsilon)
\right]\; ,
\end{align}
\end{subequations}
where for ease of notation we have defined
\be
O(\epsilon) \equiv  O \left( \frac{ {\hat S}^2 }{
      {\hat M}^2 {\hat r}^2} \right) + O \left( \frac{{\hat M}}{{\hat
        r}} \right).
\ee
Note that correction terms linear in the spin are forbidden by parity
considerations.
Computing $M$ and $r$ using Eqs.\ (\ref{eq:rdef}) yields
\be
M = {\hat M}
[  1 + O(\epsilon)], \ \ \ \ \
r = {\hat r} [  1 + O(\epsilon)].
\label{eq:rMans}
\ee
Similarly by evaluating the gradient of $r$ according to steps 3 and 4,
we find
\be
n^\alpha = {\hat n}^\alpha \left[1 + O(\epsilon) \right], \ \ \ \
y^\alpha = {\hat y}^\alpha \left[1+O(\epsilon) \right].
\label{eq:yresult}
\ee

Next, we evaluate the symmetric tensor (\ref{eq:Hdef}) using the expression
(\ref{eq:linR}) for the Riemann tensor.  The result is
\begin{eqnarray}
H_{\alpha\beta}  &=& -\frac{{\hat M}}{{\hat r}^3} (\eta_{\alpha\beta} + 3{\hat u}_\alpha
{\hat u}_\beta - {\hat n}_\alpha {\hat n}_\beta) \nonumber \\
 &&+ \frac{6}{{\hat r}^4} {\hat u}_{(\alpha}\epsilon_{\beta)\gamma\delta\epsilon}{\hat S}^\gamma
 {\hat n}^\delta {\hat u}^\epsilon\; .
\label{eq:Hans}
\end{eqnarray}
Because of the symmetries of the Riemann tensor, the tensor $H_{\alpha\beta}$ is
symmetric and has $n^\alpha$ as an eigenvector with its corresponding eigenvalue
being identically zero.
The three remaining eigenvectors at leading order in an expansion
in $1/{\hat r}$ are ${\hat u}^\alpha$,
$\epsilon_{\alpha\beta\gamma\delta} {\hat S}^\beta {\hat n}^\gamma
{\hat u}^\delta$, and a third vector that is
orthogonal to those two as well as ${\hat n}^\alpha$.
The eigenvalues associated with these eigenvectors are (again at leading order
in an expansion in $1/{\hat r}$) $2{\hat M}/{\hat r}^3$ and a repeated
eigenvalue equal to $-{\hat M}/{\hat r}^3$
for the latter two, respectively.
Therefore, if we follow step 5 and choose the normalized eigenvector
corresponding to the largest eigenvalue, we obtain
$
u^\alpha = {\hat u}^\alpha [ 1+ O(\epsilon)].
$
It follows that
\be
P^\alpha = {\hat P}^\alpha [ 1+ O(\epsilon)].
\ee

Next, from Eqs.\ (\ref{eq:linR}) and (\ref{eq:linh}), the curvature invariant
(\ref{eq:K2def}) is given by
\begin{equation}
K_2  = \frac{288{\hat M}^2}{{\hat r}^7}({\hat S}^\alpha {\hat n}_\alpha) [ 1 + O(\epsilon)]\;.
\end{equation}
Inserting this equation and the expression (\ref{eq:Hans}) for
$H_{\alpha\beta}$ into the formula (\ref{eq:Sadef}) for the intrinsic angular
momentum, we determine
\be
S^\alpha = {\hat S}^\alpha [ 1 + O(\epsilon)].
\ee
Thus, the algorithm successfully recovers the linear momentum and
intrinsic angular momentum of the spacetime.  Also, from Eqs.\
(\ref{eq:Jabdef}), (\ref{eq:hatJabdef}), and (\ref{eq:yresult}),
we find that $J^{\alpha\beta} = {\hat J}^{\alpha\beta}[1 +
O(\epsilon)]$, so that the algorithm yields the total angular momentum
of the source about the observer's location $x^\alpha$.

\subsection{Physical interpretation of the measured linear and angular momenta in
  more general contexts}
\label{subsec:interp}

In the previous subsections, we showed that an observer that is
sufficiently distant from a stationary source
of gravity can measure that source's linear and angular momentum to a
good approximation, using just the spacetime
geometry in the vicinity of the observer.
The measurement procedure required several assumptions about that
spacetime geometry: (i) the curvature invariants (\ref{eq:invariants})
needed to be positive, (ii) the vector $\nabla_a r$ needed to be
spacelike, (iii) the tensor $H_{ab}$ needed to have at least one
strictly positive eigenvalue, and (iv) the corresponding eigenvector
needed to be timelike.  These assumptions are satisfied for linearized
stationary spacetimes described by just two multipoles at sufficiently
large ${\hat r}$.  By continuity, therefore, they will also be satisfied in
regions of spacetimes that are sufficiently close to this case.
We now discuss in more detail how the measurement procedure applies
to these more general situations and spacetimes.

There are a number of physical effects that can make the spacetime
geometry measured by observers differ from the idealized case
discussed above of
asymptotic regions in linearized stationary spacetimes with two multipoles.
The effects that we consider include nonlinearities,
higher-order multipoles, nonisolated systems, and nonstationarity.
We now estimate the size of these effects in more general contexts
and thereby determine both when we might expect the assumptions listed above to break down and also when the algorithm yields physically sensible results.  
The various effects are:

\begin{itemize}

\item {\it Nonlinearities:} Our analysis above assumed that the
spacetime could be described as a linear perturbation about Minkowski
spacetime.  For an isolated, stationary source in an asymptotically
flat spacetime, there will be corrections to the metric arising from
nonlinearities.  These nonlinearities will give corrections
to the metric perturbation $h_{\alpha\beta}$ that scale\footnote{This
will be true in suitable coordinates, for which
the limit ${\hat M} \to 0$ of the metric at fixed ${\hat S}/{\hat
M}$ is the Minkowski metric in Minkowski coordinates (for example,
Cartesian Kerr-Schild coordinates in the Kerr spacetime).  For more
general coordinates (such as Boyer-Lindquist coordinates), other terms
can occur that are larger than some of the terms in Eq.\ (\ref{eq:listc})
[e.g., ${\hat S^2} / ({\hat M}^2 {\hat r}^2)$].  These larger terms are
gauge effects, and they can be ignored for the argument given here.}
as
\be
\ \ \ \ \ \ \ O\left( \frac{{\hat M}^2}{{\hat r}^2} \right), \ \ \ \ \
O\left( \frac{{\hat M} {\hat S}}{{\hat r}^3} \right), \ \ \ \ \
O\left( \frac{{\hat S}^2}{{\hat r}^4} \right).
\label{eq:listc}
\ee
The form of these corrections can be found, for example,
from the leading nonlinear terms in the post-Newtonian expansion of the
metric (given in, e.g., \cite{Blanchet2014}).
These corrections will be small\footnote{An exception is the term
$\sim {\hat M}^2/{\hat r}^2$ which will be comparable to the
$\sim {\hat S}/{\hat r}^2$ term in the metric (\ref{eq:hlinans}) when ${\hat S} \sim {\hat M}^2$.
One might expect that this term would give rise to fractional
corrections of order unity to the measured momentum and angular
momentum; the corrections, however, are suppressed, because the ${\hat
  S}/{\hat r}^2$ term is parity odd while the ${\hat M}^2 / {\hat
  r}^2$ term is parity even.}
 compared to the leading-order terms
$\sim {\hat M}/{\hat r}$ and $\sim {\hat S} / {\hat r}^2$ in
the metric perturbation (\ref{eq:hlinans}), as long as ${\hat r}$ is
large compared to ${\hat M}$, $\sqrt{\hat S}$, and ${\hat S}^{2/3} {\hat M}^{-1/3}$.
For sufficiently large ${\hat r}$, therefore, the effects of
nonlinearities can be neglected.

\item {\it Higher-order multipoles:} Our analysis in Sec.\
  \ref{subsec:AngMomDefMotivation} above included only the mass and
  spin and neglected higher-order mass and current multipoles.
However, as is well known, the effect of these multipoles will be
small at sufficiently large ${\hat r}$.  The dominant correction to
the metric perturbation in the parity-even sector will be
\be
\delta h_{\alpha\beta} \sim \frac{Q}{{\hat r}^3},
\ee
where $Q$ is the mass quadrupole.  Using the estimate $Q \sim {\hat M}
{\mathcal L}^2$, where ${\cal L}$ is the size of the source, we see that
this correction will be small compared to ${\hat M}/{\hat r}$ in the regime
\be
{\hat r} \gg {\mathcal L}.
\label{eq:asymptoticregime}
\ee
Similarly, in the parity-odd sector, the
dominant correction will be
\be
\delta h_{\alpha\beta} \sim \frac{{\mathcal S}}{{\hat r}^3},
\label{eq:here}
\ee
where ${\mathcal S} \sim {\hat S} {\mathcal L}$ is the current
quadrupole.  This correction will be small compared to the spin term in
Eq.\ (\ref{eq:hlinans}) whenever ${\hat r} \gg {\mathcal L}$.  Therefore,
in the regime (\ref{eq:asymptoticregime}), corrections to the measured linear and angular momentum
$P^\alpha$ and $J^{\alpha\beta}$ will be small.

We note that in the context of linearized gravity, it is
possible in principle to measure $P^\alpha$ and $J^{\alpha\beta}$
accurately even in the regime ${\hat r} \sim {\mathcal L}$, by
using measurement procedures more sophisticated than those envisaged
in this paper.  As is well known, in linearized gravity the charges $P^\alpha$ and $J^{\alpha\beta}$
can be extracted unambiguously from the metric perturbation using
surface integrals \cite{Misner1973}.  Therefore, a
family of observers distributed over the surface of a sphere, who make
measurements of the spacetime geometry in their vicinity and compare
notes in a suitable way, can measure $P^\alpha$ and $J^{\alpha\beta}$
with high accuracy.  In this paper, we will not need to consider
such nonlocal measurement procedures, because the issues we want to
explore are all present in the regime (\ref{eq:asymptoticregime}) in
which our local measurement procedure is sufficient.

\item {\it Nonisolated systems:}  So far we have considered observers
near isolated sources in asymptotically flat spacetimes.
Suppose, however, that there are also distant sources, or that
the spacetime is not asymptotically flat.
In linearized gravity, the effect of distant sources can be quantified in terms of the
tidal tensor ${\cal E}_{ij}$ (the electric components of the
associated Riemann tensor).  The corresponding fractional corrections to the
linear momentum measured by observers using the procedure
of Sec.\ \ref{subsec:AngMomStat} will be of order
$
\sim {\cal E} {\hat r}^3/{\hat M}.
$
Similarly the fractional corrections to the angular momentum will be
of order $
\sim {\cal B} {\hat r}^4/{\hat S},
$
where ${\cal B}_{ij}$ is the magnetic tidal tensor.
These effects limit the accuracy and utility of our measurement method of
Sec.\ \ref{subsec:AngMomStat} above.
Within the context of linearized gravity, it is possible to circumvent
this difficulty using the nonlocal measurement method discussed above,
which uses the angular dependence to disentangle the effects of the locally
produced curvature $\sim {\hat M} / {\hat r}^3$ from the curvature
 ${\cal E}_{ij}$ produced by distant sources.

When nonlinearities are included, however, there is
an unavoidable ambiguity: the linear and angular momenta of
individual objects cannot be defined in general.  We can
estimate the ambiguities from nonlinearities using the fact that
different definitions of the ``mass of an object'' in post-1-Newtonian
theory differ by a
quantity of order the tidal-interaction energy, $Q_{ij} {\cal  E}_{ij}$,
where $Q_{ij} \sim {\hat M} {\cal L}^2$ is a mass quadrupole.
Therefore, objects of mass ${\hat M}$, size ${\cal L}$ and separated by
distances $\sim {\cal D}$ have an uncertainty or ambiguity in their
masses of order\footnote{This estimate is valid for generic sources which have a nonvanishing intrinsic quadrupole moment.  It is not valid for spherically symmetric sources whose intrinsic quadrupole vanishes.  For such sources, the scaling of the mass ambiguity can be estimated from the quadrupoles $Q \sim {\hat M} {\cal L}^5 / {\cal D}^3$
induced by tidal interactions; it is of order  $
\Delta {\hat M}/{\hat M} \sim {\hat M} {\cal L}^5 /
{\cal D}^6$.}
\be
\Delta {\hat M}/{\hat M} \sim {\hat M} {\cal L}^2 /
{\cal D}^3 \; .
\ee
The measurement method discussed in Sec.\ \ref{subsec:AngMomStat} above will
be subject to this ambiguity; however, in many situations the ambiguity will
be negligible.

\item {\it Nonstationary systems:} For dynamical, radiating sources,
it is immediately clear that our measurement procedure will not be
applicable in general.  The reason is that the Weyl tensor for
radiated gravitational waves falls off at large ${\hat r}$ as $1/{\hat
r}$, whereas the static piece of the Weyl tensor associated with the
mass and spin falls off as $1/{\hat r}^3$.  Therefore, at sufficiently
large ${\hat r}$, if an observer measures the Riemann tensor and its
derivatives at her location, her result will be dominated by the radiative
pieces of the metric, and the measurement method of Sec.\
\ref{subsec:AngMomStat} above will fail.

As discussed in the Introduction, however, the measurement method can
still yield interesting information about dynamical systems, for
an {\it intermittently stationary} spacetime (by which we mean
a spacetime which is stationary at early times and again at late
times).  Observers can apply the measurement procedure at early and at
late times and then attempt to compare their results.  This scenario is
discussed in detail in the remaining sections of the paper.

As an aside, we note that we can classify nonstationary systems into
two types.  The first is what we will call {\it asymptotically linear}
systems, that is, systems for which the linear approximation is
valid\footnote{Here the assumption is that the linear approximation is
  valid in a neighborhood of some two-sphere which encloses the
  source, not the weaker assumption the linear approximation is valid in a neighborhood of
  some observer.}
at sufficiently large ${\hat r}$.  For these systems, one can define
unambiguous linear and angular momenta using surface integrals, and
they can be measured using the nonlocal measurement procedure
discussed above.  Our local measurement procedure can work for such
systems, but only if ${\cal L} \ll {\hat r} \ll \lambda$,
where $\lambda$ is the wavelength of the radiation.
The second type of system, {\it asymptotically nonlinear} systems, are
those for which the linear approximation is not valid at large
${\hat r}$.  These are the systems for which the BMS asymptotic
symmetry group is most relevant.  Neither our local measurement procedure
nor the nonlocal measurement procedure based on surface integrals of
linearized theory apply to systems in this regime.\footnote{Except to
the extent that measurements before and after the nonstationarity
can probe effects of the nonstationarity, as we discuss in the
remainder of this paper.}

\end{itemize}

\subsection{Nonuniqueness of the measurement algorithm}
\label{subsec:nonunique}

The algorithm discussed above is not uniquely determined by the requirement that it give the correct answer in linearized stationary spacetimes with two multipoles, because the information about the linear and angular momentum of the spacetime is encoded redundantly in the Riemann tensor and its first two derivatives at any point.
Therefore, there are several methods that can be used to extract these momenta.
For example, Eq.\ (\ref{eq:ydef}) could be replaced by
$y^a = - \nabla^a r^2/2$, which would give the same result to leading order.

In stationary linearized spacetimes with two multipoles, there
is a unique and accepted definition of the linear and angular momentum of the
spacetime; therefore, any nonuniqueness or ambiguities in the measurement
procedure must vanish in this limit as the measurement is taken at large
distances from the source.
More specifically, this implies that
the effects of these ambiguities all scale as $1/r$ as $r \to \infty$ (or as $1/v$, where $v$ is a null coordinate with goes to infinity at future null infinity).  Most importantly, they are small compared to the observer
dependence of angular momentum that we discuss in the remainder of the paper (that characterized by generalized holonomies, which we show gives rise to finite effects in the limit $v \to \infty$).

\subsection{Accuracy of measurement of the center-of-mass worldline}
\label{subsec:AngMomCOM}

The procedure discussed above allows an observer to measure the
angular momentum of the spacetime about his own location to an
accuracy of $\epsilon = {\hat M}/{\hat r}$:
\be
J^{\alpha\beta} = {\hat J}^{\alpha\beta} \left[ 1 + O(\epsilon)
\right].
\ee
In particular, the displacement vector $y^\alpha$ from the observer to
the center-of-mass worldline [cf.\ Eq.\ (\ref{eq:Jabdef}) above]
will be measured with this accuracy:
\be
y^{\alpha} = {\hat y}^{\alpha} \left[ 1 + O\left(\frac{{\hat M} }{\hat
      r} \right) \right].
\ee
However, ${\hat y}^\alpha$ is of order ${\hat r}$, and, therefore, the
error in the measurement is of order
\be
\delta y^{\alpha} \sim {\hat M}.
\ee
This error is large; it is of the same order as the maximum displacements caused by
gravitational-wave memory effects.\footnote{It is possible, however, to modify the measurement method to increase the
accuracy as follows.
Modify the definitions of $M$ and $r$ in Eqs.\ (\ref{eq:rdef}) to
\begin{subequations}
\begin{eqnarray}
M &=& \frac{15 \sqrt{5} K_1^2}{4\mathcal K_1^{3/2}}
 \left[1 - \frac{15\sqrt 3 (K_1)^{3/2}}{4\mathcal K_1} \right]
\;,\\
r &=& \sqrt{\frac{15 K_1}{\mathcal K_1}}
 \left[1 - \frac{5\sqrt 3 (K_1)^{3/2}}{4\mathcal K_1} \right]
\; ,
\end{eqnarray}
\label{eq:rdef1}%
\end{subequations}
and leave the rest of the measurement algorithm unaltered.
Then the fractional error in measurement of $y^\alpha$ is decreased to
$O({\hat M}^2/{\hat r}^2)$, and the errors in $y^\alpha$ vanish as the
observers approach future null infinity.
This modified algorithm is derived from the expressions for the curvature
invariants of the Kerr spacetime in Boyer-Lindquist coordinates and therefore
yields the Boyer-Lindquist radial-coordinate values at the observers' 
locations.}

\section{Affine transport and generalized holonomy: Properties and
application to angular momentum}
\label{sec:GenHolDef}

We now turn to the question of how two observers at different locations in a curved spacetime can compare values of linear and angular momentum.
The philosophy we adopt is to imagine that the observers attempt to compare values using the same methods they would use in special relativity (i.e.,
in the absence of gravity).

The first part of this section introduces a curve-dependent transport law,
which we call affine transport and which serves as the basis for our method
of comparing angular momentum.
The next subsection describes how the affine transport can be used to
compare values of the angular momentum defined at different spacetime points.
The final subsection describes the affine transport around a closed curve,
which we call the generalized holonomy, and it explains its relation
to the inevitable observer dependence of angular momentum in curved spacetimes.

\subsection{Definition of an affine transport law}
\label{subsec:AffineTrans}

In this section, we define a transport law that can be used
to transport vectors along curves and which is a generalization of parallel
transport.
Let $\mathcal C$ be a curve between the spacetime points $\mathcal P$ and
$\mathcal Q$, and let the curve have tangent vector $\vec k$.
Next, define a map $\chi_{\mathcal C}$ from $T_{\mathcal P}(M)$ to
$T_{\mathcal Q}(M)$ through the solution of the differential equation
\begin{equation}
\label{eq:GenHolDef}
\nabla_{\vec k} \vec \xi = \alpha \vec k \; .
\end{equation}
Here $\alpha$ is a dimensionless constant.
Namely, starting from an initial condition $\vec \xi_{\mathcal P}$ in
$T_{\mathcal P}(M)$, we solve the differential equation to obtain the value $\vec \xi_{\mathcal Q}$ of $\vec \xi$ at ${\mathcal Q}$.
The image of $\vec \xi_{\mathcal P}$ under
the map $\chi_{\mathcal C}$ is then defined to be
$\vec \xi_{\mathcal Q}$.
Since we are not aware of a name for this specific transport law, we will
call it the {\it affine transport} of the vector $\vec \xi$ along
the curve $\mathcal C$ with tangent $\vec k$.
This map satisfies six important properties that are listed below:

\begin{enumerate}

\item It is independent of the choice of parametrization along the curve
(which follows because both sides of the equation are linear in the tangent
to the curve $\vec k$).

\item When two curves are composed, the composition of maps is equivalent
to the map on the composed curve (i.e., if
$\mathcal C = \mathcal C_1 \cup \mathcal C_2$ then
$\chi_{\mathcal C} = \chi_{\mathcal C_1} \circ \chi_{\mathcal C_2}$).

\item For a fixed curve, $\mathcal C$, Eq.\ \eqref{eq:GenHolDef} is a linear
differential equation in $\vec \xi$.  The solution for given initial data,
therefore, can be expressed as the sum of two terms: the first term is the solution of the homogeneous differential equation (parallel transport) with the same initial data, and the second is the solution of the inhomogeneous differential equation with zero initial data.
The complete solution is
\be
\xi_{\mathcal Q}^{\bar a} =
\Lambda_{\mathcal{PQ}}{}^{\bar a}{}_a \xi_{\mathcal P}^a +
\alpha \Delta \xi^{\bar a}_{\mathcal{PQ}} \; ,
\label{eq:ATdef}
\ee
where
$\Lambda_{\mathcal{PQ}}{}^{\bar a}{}_a$ denotes the parallel transport
operation from $\mathcal P$ to $\mathcal Q$ and
$\Delta\xi^{\bar a}_{\mathcal{PQ}}$ is the inhomogeneous solution for $\alpha=1$.
The notation here is that overlined indices are associated with the point
$\mathcal Q$ and indices without extra adornment are associated with
$\mathcal P$.

\item It follows that (unlike parallel transport) affine transport
does {\it not} preserve the norm of the transported vector.

\item For geodesic curves, one can show that the inhomogeneous part
of the solution $\Delta\vec\xi_{\mathcal{PQ}}$ is just the tangent to the
curve at the point $\mathcal Q$ (i.e., $\alpha \vec k_{\mathcal Q}$).
\footnote{The calculation which shows this is short:
Let $\lambda\in[0,1]$ be an affine parameter along a geodesic curve
with $\lambda=0$ corresponding to $\mathcal P$ and $\lambda=1$ be
the point $\mathcal Q$, and denote the directional derivative along the
geodesic by $\nabla_{\vec k} = D/D\lambda$.
For a point $\mathcal P'$ between $\mathcal P$ and $\mathcal Q$, one can
confirm that
$\xi_{\mathcal P'}^{\bar a} = \Lambda_{\mathcal{PP'}}{}^{\bar a}{}_a
\xi_{\mathcal P}^a + \alpha \lambda k^{\bar a}_{\mathcal P'}$
is the solution along the geodesic curve, because
$\nabla_{\vec k} \Lambda_{\mathcal{PP'}}{}^{\bar a}{}_a \xi_{\mathcal P}^a = 0$
and $\nabla_{\vec k} \vec k = 0$.
Evaluating the expression at $\lambda=1$, one finds that the inhomogeneous
part of the solution is $\alpha \vec k_{\mathcal Q}$.}

\item Finally, for curves in a flat spacetime,
$\Delta\vec\xi_{\mathcal{PQ}}$ is just the vectorial displacement
$\vec {\cal Q} - \vec {\cal P}$ in any inertial coordinate system.
In particular, it vanishes for closed curves in a flat spacetime,
because, as we show later, it is only nontrivial in the
presence of spacetime curvature.

\end{enumerate}

We will use these properties frequently in the calculations in
the remainder of this paper.

\subsection{Application to a curve-dependent definition of angular-momentum transport}
\label{subsec:GenHolApps}

Using the affine transport law, we can define a method of comparing the
local values of angular momentum at two different spacetime points, by
transporting angular momenta from one spacetime point to another, in a
curve-dependent manner.

We define a map from
$\mathcal G_{\mathcal P}^*$ to $\mathcal G_{\mathcal Q}^*$ that depends on
a choice of curve $\mathcal C$ that joins these two points.
Because the elements of $\mathcal G_{\mathcal P}^*$ act on maps in
$\mathcal G_{\mathcal P}$ [those from the tangent space $T_{\mathcal P}(M)$
to itself], a natural map is one based on the affine transport of elements in
$\mathcal G_{\mathcal P}$ to elements of $\mathcal G_{\mathcal Q}$.
Namely, for $h_{\mathcal P} \in \mathcal G_{\mathcal P}$, the corresponding
$h_{\mathcal Q} \in \mathcal G_{\mathcal Q}$ is defined by
\begin{equation}
h_{\mathcal P} = \chi_{\mathcal C}^{-1} \circ h_{\mathcal Q} \circ
\chi_{\mathcal C} \, ,
\label{eq:AlgTrans}
\end{equation}
where $\chi_{\mathcal C}$ is the affine transport
along $\mathcal C$.
For an element $q_{\mathcal P} \in \mathcal G_{\mathcal P}^*$, therefore
we define the corresponding element
$q_{\mathcal Q} \in \mathcal G_{\mathcal Q}^*$ by
\begin{equation}
q_{\mathcal Q}(h_{\mathcal Q} ) = q_{\mathcal P}
(\chi_{\mathcal C}^{-1} \circ h_{\mathcal Q} \circ \chi_{\mathcal C})
= q_{\mathcal P}(h_{\mathcal P}) \; .
\label{eq:AngMomTransform}
\end{equation}
To recover the correct transformation properties of angular momentum
under displacements with this definition, it is necessary to choose the value
\be
\alpha = -1
\ee
of the parameter in the definition (\ref{eq:GenHolDef}) of the function
$\chi_{\mathcal C}$, which we now show by writing this mapping from the
angular-momentum space $\mathcal G_{\mathcal P}^*$ to the angular-momentum
space $\mathcal G_{\mathcal Q}^*$ in a more explicit notation.

To do so, let us represent elements of the algebra
$h_{\mathcal P}$ as pairs
\begin{equation}
h_{\mathcal P} \leftrightarrow
(\kappa_a^{\mathcal P}, \omega_{ab}^{\mathcal P})
\end{equation}
and the map between algebras at different points $\mathcal P$ and
$\mathcal Q$, $\chi_{\mathcal C}$, as
\begin{equation}
\chi_{\mathcal C} \leftrightarrow
(\alpha\Delta\xi_{\mathcal{PQ}}^{\bar a},
{{\Lambda_{\mathcal{PQ}}}^{\bar a}}_a) \, ,
\end{equation}
where the quantities in this equation are exactly those appearing in Eq.\
\eqref{eq:ATdef}.
The transformation rule for the algebra elements \eqref{eq:AlgTrans}
then has the representation in terms of these pairs as
\begin{align}
& (\kappa_{a}^{\mathcal P},\omega_{ab}^{\mathcal P}) =
\nonumber \\
& ({{\Lambda_{\mathcal P\mathcal Q}}^{\bar a}}_a \kappa^{\mathcal Q}_{\bar a}
+ \alpha \Lambda_{\mathcal{QP}}{}^{\bar a}{}_a
\Lambda_{\mathcal{QP}}{}^{\bar b}{}_b \omega_{\bar a\bar b}^{\mathcal Q}
\Delta\xi_{\mathcal{PQ}}^b, \Lambda_{\mathcal{QP}}{}^{\bar a}{}_a
\Lambda_{\mathcal{QP}}{}^{\bar b}{}_b \omega_{\bar a\bar b}^{\mathcal Q}) \, ,
\label{eq:AlgTransPairs}
\end{align}
which respects the multiplication rule for a semidirect-product structure.
In the above expression, we have used the notation
$\Delta\xi^a_{\mathcal{PQ}} = \Lambda_{\mathcal{PQ}}{}_{\bar a}{}^a
\Delta\xi^{\bar a}_{\mathcal{PQ}}$ and the fact that
$\Lambda_{\mathcal{QP}}{}^{\bar a}{}_a$ and
$\Lambda_{\mathcal{PQ}}{}^{\bar a}{}_a$ are related in accord with the typical
notation for the inverse of the parallel propagator.
Keeping in mind the representation of the maps $q_{\mathcal P}$ as
\begin{equation}
q_{\mathcal P}(h_{\mathcal P}) = P^a_{\mathcal P} \kappa_a^{\mathcal P}
- \frac 12 J^{ab}_{\mathcal P} \omega_{ab}^{\mathcal P}
\end{equation}
given in (\ref{eq:eldef}), we can then take the definition of the
transformation property of angular momentum in Eq.\ \eqref{eq:AngMomTransform}
above and substitute in the result of \eqref{eq:AlgTransPairs}; by equating
the coefficients of $\omega_{\bar a\bar b}^{\mathcal Q}$ and
$\kappa_{\bar a}^{\mathcal Q}$, we find that
the angular momenta at the two points are related by
\begin{equation}
J_{\mathcal Q}^{\bar a\bar b} =  \Lambda_{\mathcal{QP}}{}^{\bar a}{}_a
\Lambda_{\mathcal{QP}}{}^{\bar b}{}_b \left(J_{\mathcal P}^{ab}
+ 2 \alpha \Delta\xi^{[a}_{\mathcal{PQ}} P_{\mathcal P}^{b]}\right) \; ,
\label{eq:AngMomObsDepAlpha}
\end{equation}
and the corresponding momenta are related by
\be
P^{\bar a}_{\mathcal Q} = \Lambda_{\mathcal{PQ}}{}^{\bar a}{}_a
P^a_{\mathcal P} \, .
\label{eq:LinMomObsDep}
\ee
We, therefore, see that to have the usual transformation law for angular
momentum
\begin{equation}
J_{\mathcal Q}^{\bar a\bar b} =  \Lambda_{\mathcal{QP}}{}^{\bar a}{}_a
\Lambda_{\mathcal{QP}}{}^{\bar b}{}_b \left(J_{\mathcal P}^{ab}
- 2\Delta\xi^{[a}_{\mathcal{PQ}} P_{\mathcal P}^{b]}\right) \; ,
\label{eq:AngMomObsDep}
\end{equation}
we must choose $\alpha=-1$.
An alternative and simpler formulation of the transport law given by
Eqs.\ (\ref{eq:AngMomObsDep})
and (\ref{eq:LinMomObsDep}) is discussed in Appendix \ref{sec:alternative}.

If we decompose the angular momentum $J^{ab}$ into an intrinsic spin
$S^a$ and a displacement vector $y^a$ using the definitions
\begin{subequations}
\begin{align}
y^a = -\frac 1{M^2} J^{ab} P_b \, , \\
S^a = \frac 1{2M} {\epsilon_{bcd}}^a P^b J^{cd} \, ,
\end{align}%
\end{subequations}
then from Eqs.\ \eqref{eq:LinMomObsDep} and \eqref{eq:AngMomObsDep}, the
fact that $P^a P_a = -M^2$, and that $y^a P_a = 0$, we can show after
some algebra that the spin is parallel transported just like the linear 
momentum,
\be
S^{\bar a}_{\mathcal Q} = \Lambda_{\mathcal{PQ}}{}^{\bar a}{}_a
S^a_{\mathcal P} \, ,
\label{eq:SpinObsDep}
\ee
while the displacement vector transforms as
\be
y_{\mathcal Q}^{\bar a} =  \Lambda_{\mathcal{QP}}{}^{\bar a}{}_a
\left(y_{\mathcal P}^{b}
- \Delta\xi^{a}_{\mathcal{PQ}}\right) \; .
\ee

Additional properties of the affine transport for closed curves are
discussed next.

\subsection{Generalized holonomy: A measure of observer dependence of angular momentum}
\label{subsec:GenHol}

For closed curves starting from a point $\mathcal P$, the affine transport around the curve defines a {\it generalized holonomy}, a map from the tangent space at $\mathcal P$ to itself.  For flat spacetimes, the generalized holonomy is always the identity map.
Specializing the result \eqref{eq:AngMomObsDep} to closed curves (when
$\mathcal Q$ is the same point as $\mathcal P$), yields the mapping
\begin{equation}
J^{ab} \rightarrow \Lambda{}^a{}_c \Lambda^{b}{}_d
(J^{cd} - 2 \Delta \xi^{[c} P^{d]})\; .
\label{eq:AngMomGenHol}
\end{equation}
Thus, if there is a nontrivial holonomy of parallel transport
or a nonzero inhomogeneous solution, then observers along the curve will
find that angular momentum is observer dependent.
The extent to which a generalized holonomy is nontrivial is a measure
of how much spacetime curvature is an obstruction to separated observers
arriving at a consistent definition of angular momentum.

As a simple example of this generalized holonomy, consider
an infinitesimal quadrilateral starting from a point $\mathcal P$ with
legs given by $\epsilon u^a$ and $\epsilon v^a$ where $\epsilon$ is small.
The quadrilateral is traversed first in the direction of $u^a$, then $v^a$, 
then $-u^a$, then $-v^a$.
If we start at $\mathcal P$ with some initial vector $\xi^a$ and solve the
transport equation (\ref{eq:GenHolDef}) around the loop,
a relatively straightforward calculation shows that the homogeneous part of
the solution is
\be
\xi^a + \epsilon^2 R^a{}_{bcd}\xi^b v^c u^d + O(\epsilon^3).
\ee
This is the usual expression for the holonomy around a small loop.
The inhomogeneous part of the solution is
\be
\Delta \xi^a = \tfrac 12 \epsilon^3
R^a{}_{bcd} v^c u^d (u^b + v^b) + O(\epsilon^4).
\ee
A more detailed calculation is given in Ref.\ \cite{Vines2014}.
Thus, while the holonomy of parallel transport is proportional to the Riemann
tensor contracted with the area of the quadrilateral, the generalized holonomy
contains an additional term proportional to the Riemann tensor contracted with
both the area and the perimeter of the region.

\subsection{Relation between generalized holonomy and gravitational-wave memory}
\label{subsec:GenHolFermi}

In this section, we give a precise and covariant definition of an observable
that can be interpreted as a ``gravitational-wave memory'' generalized to an arbitrary spacetime.
We then show that the generalized holonomy around a suitably constructed loop contains
information about this covariant gravitational-wave memory, showing a very general relationship
between these two observables.

However, we also show that the generalized holonomy contains additional
information and specifically contains three other independent pieces,
each of which could arise as a kind of ``memory'' effect due to the
passage of a burst of gravitational waves.  The first is a
difference in proper time measured by two observers (a gravitational
redshift effect).  The second is a relative boost of two initially
comoving observers.  The third is a relative rotation of the inertial
frames of two observers.
In the limit of nearby geodesics at large distances from
a source emitting a burst of gravitational waves with memory, these effects
reduce to a combination of more familiar notions of gravitational-wave memory
arising from solutions of the equation of geodesic deviation and of
differential frame dragging and the difference in proper time of nearby 
geodesics \cite{Flanagan2015}.
(We are also investigating the relationship between the new gravitational-wave
memory of Pasterski {\it et al}., \cite{Pasterski2015} and the memory
effects quantified by the generalized holonomy in \cite{Flanagan2015}.)
We now describe a calculation that elucidates the relationship between
the generalized holonomy and the ordinary memory plus differences in
proper time, relative rotations, and relative boosts.

Consider two freely falling observers $A$ and $B$ in an arbitrary
spacetime.  We fix attention on an interval of $A$'s worldline between
two events $\mathcal P$ and $\mathcal R$, where $A$'s proper time
$\tau$ varies between $\tau_1$ and $\tau_2$, as illustrated in Fig.\
\ref{fig:Observers}.
We denote by ${\vec u}_A(\tau)$ the 4-velocity of $A$ along her worldline.
We also introduce an orthonormal
tetrad ${\vec e}_{\hat \alpha}(\tau)$ which is parallel transported
along $A$'s worldline, where ${\vec e}_{\hat 0} = {\vec u}_A$.

At each point on $A$'s worldline, there is a unique spatial vector
\be
\xi_B^{\hat i}(\tau) {\vec e}_{\hat i}(\tau)
\label{eq:initialvector}
\ee
such that the exponential map evaluated on this vector is a point $z_B(\tau)$
on $B$'s worldline.\footnote{Uniqueness requires that $B$ is
  sufficiently close to $A$ to be inside a convex normal
  neighborhood.} Or, equivalently, $(\tau, \xi^{\hat i}_B(\tau))$
gives the location of $B$'s worldline in
Fermi normal coordinates centered on $A$'s worldline.
We denote by ${\vec u}_B(\tau)$ the 4-velocity of observer $B$ at
the point $z_B(\tau)$.
We denote by $\mathcal Q$ and $\mathcal S$ the initial and final points
$z_B(\tau_1)$
and $z_B(\tau_2)$ on B's worldline (see Fig.\
\ref{fig:Observers} below).
Finally, we let ${\vec f}_{\hat \alpha}(\tau)$ be the orthonormal
tetrad at $z_B(\tau)$ obtained by parallel transporting ${\vec
  e}_{\hat \alpha}(\tau)$ from the corresponding point on $A$'s
worldline along the spatial geodesic with initial tangent (\ref{eq:initialvector}).\footnote{Note that the parameter $\tau$ need not be the proper time along 
$B$'s worldline, and the orthonormal tetrad ${\vec f}_{\hat \alpha}$ need not 
be parallel transported along $B$'s worldline.}

We assume that the observers $A$ and $B$ are initially comoving, in
the sense that ${\vec f}_{\hat 0}(\tau_1)$ is $B$'s 4-velocity at
$\mathcal Q$.
We define a closed loop $\mathcal C$ by starting at $\mathcal R$,
traveling along $A$'s worldline back to $\mathcal P$, traveling
along the spatial geodesic with initial tangent
(\ref{eq:initialvector}) to $\mathcal Q$, traveling along $B$'s
worldline to $\mathcal S$, and then back to $\mathcal R$
along the spatial geodesic
whose final tangent at $\mathcal R$ is the vector
(\ref{eq:initialvector}) at $\tau = \tau_2$.

The inhomogeneous part of the generalized holonomy about the loop
$\mathcal C$ is given by
\begin{eqnarray}
{\vec {\Delta \xi}} = \left[ \xi^{\hat i}_B(\tau_1) - \xi^{\hat
    i}_B(\tau_2) \right] {\vec e}_{\hat i} + (\Delta \tau_B - \Delta
\tau_A) {\vec w}_B \nonumber \\
\mbox{} + \xi^{\hat i}_B(\tau_1) \left( {\bf \Lambda} \cdot {\vec
    e}_{\hat i} - {\vec e}_{\hat i} \right).\ \ \ \
\label{eq:generalans}
\end{eqnarray}
Here $\Delta \tau_A = \tau_2 - \tau_1$ is the interval of $A$'s proper
time between $\mathcal P$ and $\mathcal R$, and $\Delta \tau_B$ is the
interval of $B$'s proper time between $\mathcal Q$ and $\mathcal S$.
The quantity $\Lambda^a_{\ b}$ is the usual holonomy around the loop $\mathcal C$.
Finally ${\vec w}_B$ is the 4-vector at ${\cal R}$ obtained by
parallel transporting $B$'s 4-velocity ${\vec u}_B(\tau_2)$ at $\mathcal S$ along the
spatial geodesic to $\mathcal R$.  Equivalently, it can be obtained by
acting with the holonomy around the loop on $A$'s 4-velocity at
$\mathcal R$,  $w_B^a =
\Lambda^a_{\ b} u_A^b(\tau_2)$.

The first term in the generalized holonomy (\ref{eq:generalans}) can
be interpreted as (a generalization of) the gravitational-wave memory effect.  
It is the change in the relative displacement of the observers $A$ and $B$, as
seen by $A$ in her Fermi normal coordinates, when $A$ and $B$ are
initially comoving.
The second term depends on the difference in the proper times measured
by $A$ and $B$ along the corresponding segments of their worldlines.
It also depends on the boost that relates the final velocity of $B$ to
that of $A$.  Finally, the third term depends on the holonomy
$\Lambda^a_{\ b}$ around the loop, which in general will consist of a
spatial rotation together with the aforementioned boost.

We now turn to the derivation of the formula (\ref{eq:generalans}).
The inhomogeneous part of the generalized holonomy can be obtained by
solving the differential equation (\ref{eq:GenHolDef}) with $\alpha =1$ around the
loop $\mathcal C$ starting with $\vec \xi_{\mathcal R}=0$ at the initial point
$\mathcal R$.  The solution at the next point $\mathcal P$ can be
obtained from the fifth property listed in Sec.\
\ref{subsec:AffineTrans} above,
that the inhomogeneous term for a geodesic is just the tangent to the geodesic.
It is given by
\be
{\vec \xi}_{\mathcal P} = - \Delta \tau_A {\vec u}_A(\tau_1).
\label{eq:ic}
\ee
We now solve the differential equation along the leg $\mathcal P
\mathcal Q$ of the loop.  The solution at $\mathcal Q$ will be the sum
of the parallel transport of the initial condition (\ref{eq:ic}),
together with an inhomogeneous term that is the tangent to the spatial
geodesic.  The result is
\be
{\vec \xi}_{\mathcal Q} = - \Delta \tau_A {\vec u}_B(\tau_1) +
\xi^{\hat i}_B(\tau_1) {\vec f}_{\hat i}(\tau_1).
\label{eq:ic1}
\ee
Here the first term is the parallel transport term, and we have used
the fact that the parallel transport of $A$'s initial 4-velocity is
$B$'s initial 4-velocity.  The second term is the tangent to the
spatial geodesic at $\mathcal Q$, from the definitions (\ref{eq:initialvector}) of
$\xi^{\hat i}_B$ and of ${\vec f}_{\hat \alpha}$.

Next, we solve the differential equation along the segment $\mathcal Q
\mathcal S$ of $B$'s worldline.  Since $B$ parallel transports his own
4-velocity, the result is
\be
{\vec \xi}_{\mathcal S} = - \Delta \tau_A {\vec u}_B(\tau_2) +
\xi^{\hat i}_B(\tau_1) {\bf \Gamma} \cdot {\vec f}_{\hat i}(\tau_1) +
\Delta \tau_B {\vec u}_B(\tau_2).
\label{eq:ic2}
\ee
Here $\Gamma^a_{\ b}$ is the parallel transport operator from
$\mathcal Q$ to $\mathcal S$, and the last term is the inhomogeneous
term, the tangent to $B$'s worldline at $\mathcal S$.
Finally we transport this result along the leg $\mathcal S \mathcal R$
of the loop.
When we parallel transport $B$'s 4-velocity ${\vec u}_B$, the
result is the vector ${\vec w}_B$ defined above.  Similarly, when we
parallel transport the vector ${\bf \Gamma} \cdot {\vec f}_{\hat i}$,
the result is the holonomy operator $\Lambda^a_{\ b}$ of the loop acting on the basis
vector ${\vec e}_{\hat i}$ at $\mathcal R$.  This is because ${\vec
  e}_{\hat i}$ is parallel transported along $\mathcal R \mathcal P$,
and because ${\vec f}_{\hat i}$ is obtained from ${\vec e}_{\hat i}$
by parallel transporting along $\mathcal P \mathcal Q$.
Thus, we obtain
\begin{eqnarray}
{\vec \xi}_{\mathcal R} =
(\Delta \tau_B - \Delta
\tau_A) {\vec w}_B
+ \xi^{\hat i}_B(\tau_1) {\bf \Lambda} \cdot {\vec e}_{\hat i}
- \xi^{\hat i}_B(\tau_2) {\vec e}_{\hat i} ,\ \ \ \
\nonumber \\
\mbox{}
\label{eq:generalans1}
\end{eqnarray}
where the last term is the inhomogeneous term.
This is equivalent to the formula (\ref{eq:generalans}).

\section{Generalized holonomy in linearized gravity}
\label{sec:GenHolLin}

\begin{figure}
\includegraphics[width=\columnwidth]{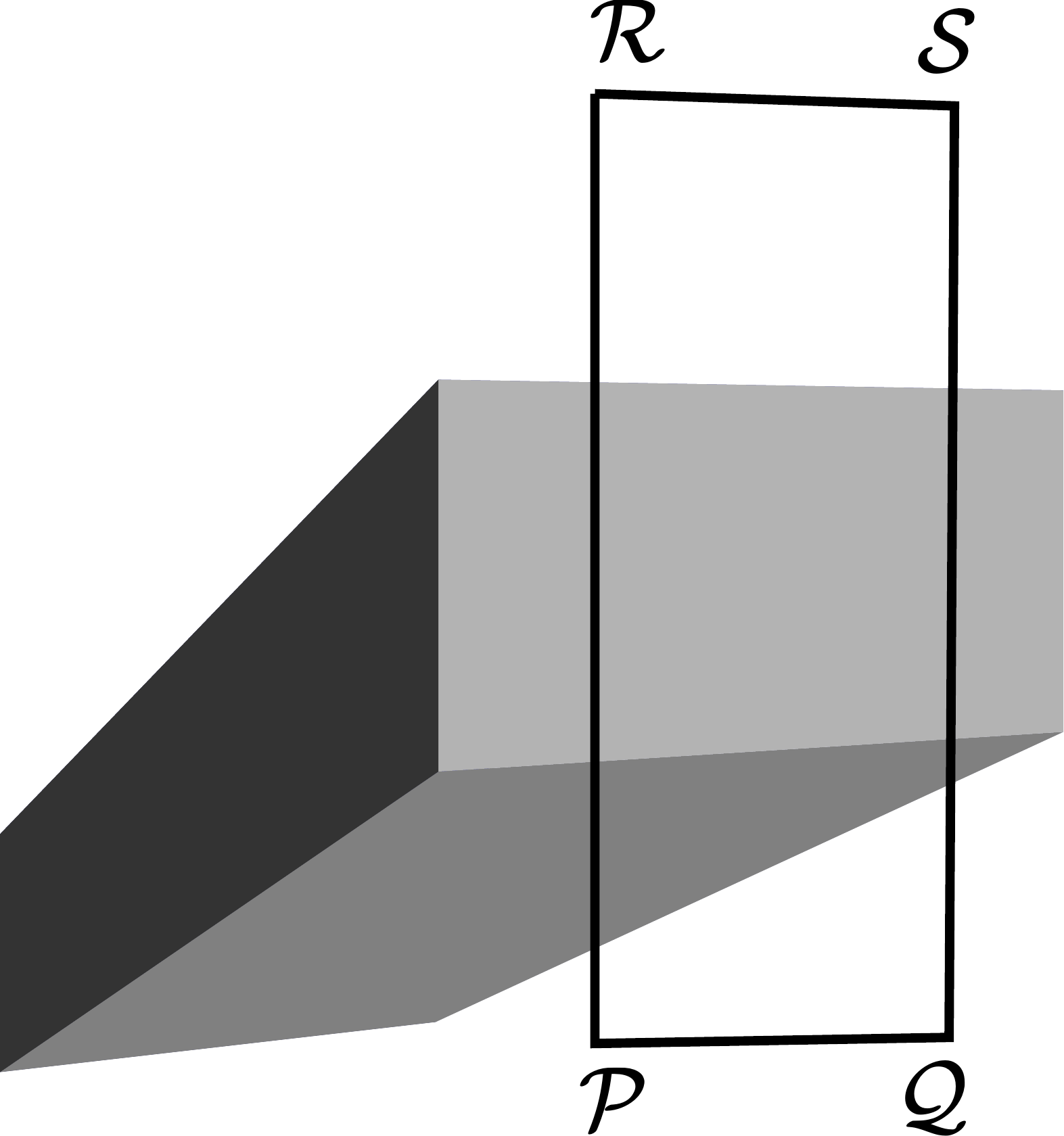}
\caption{Spacetime diagram of a burst of gravitational waves and the
curve used to compute the generalized holonomy.
The gray region represents the spacetime location of the gravitational
waves, while the unshaded regions are Minkowski spacetimes before and
after the burst.
The curve bounded by $\mathcal P$ and $\mathcal R$ is the worldline of
observer $A$, and that bordered by $\mathcal Q$ and $\mathcal S$ is that
of $B$.
The curves with endpoints $(\mathcal P,\mathcal Q)$ and
$(\mathcal R, \mathcal S)$ are spacelike geodesics before and after the
burst, respectively, which are just straight lines in the flat spacetime
regions.}
\label{fig:Observers}
\end{figure}

This section provides two related examples of the generalized holonomy.
Both spacetimes consist of a flat Minkowski region followed by a burst of
gravitational waves with memory, after which the spacetimes settle to a
different Minkowski region.
The first example treats a linearized plane wave, which reproduces the
result of Sec.\ \ref{subsec:NewtArg} in a covariant language.
The second example deals with a linearized pulse of waves heading radially
outward from a pointlike source.
This more general example gives an indication of the magnitude and the form
of the disagreement that observers will have when measuring angular momentum.

A schematic spacetime diagram with the curve used to compute the generalized
holonomy is depicted in Fig.\ \ref{fig:Observers}.  As discussed in the previous section,
there are two freely falling observers $A$ and $B$, and we consider a closed curve
consisting of segments of their worldlines together with spatial geodesics that join
the two worldlines.  In this section, we additionally assume that spacetime is flat
initially and at late times (the unshaded portions of the diagram), and that at intermediate
times there is a burst of gravitational waves present (the gray shaded region).
The two spacelike curves---the one, before the burst, with endpoints at 
$\mathcal P$ and $\mathcal Q$ and the other, after the burst, with endpoints 
at $\mathcal R$ and $\mathcal S$---are geodesics of Minkowski space (i.e., 
straight lines in some surface of constant time).
The figure describes both the plane wave (Sec. \ref{subsec:GenHolNewt}) and a local region of the
radially propagating gravitational wave (Sec. \ref{subsec:GenHolGen}).

\subsection{Generalized holonomy for a gravitational plane wave with memory}
\label{subsec:GenHolNewt}

We consider a spacetime which is flat at early and at late times
and which contains a linearized plane wave at intermediate times.
We use the conventions described in Sec.\ \ref{subsec:NewtArg} above:
the metric is given by the expression (\ref{metric00}) in global
TT coordinates $(T,X^i)$, with the metric perturbation being $h_{ij}(T-Z)$.
This metric perturbation vanishes at early times but at late times
asymptotes to the constant value $h_{ij}^\infty$.
We also introduce a coordinate system $(t,x^i)$ which at early times
is an inertial coordinate system and coincides with the TT coordinates
$(T,X^i)$ and which at late times is again inertial and related to
the TT coordinates by Eq.\ (\ref{inertialcoords}).

The observers $A$ and $B$ are freely falling and are therefore stationary
with respect to the TT coordinates, with $X^i = X^i_A =
$ constant for $A$, and $X^i = X^i_B = $ constant for
$B$.  The inertial-frame locations of the observers at early times are
\be
x^i_A = X^i_A, \ \ \ \ x^i_B = X^i_B,
\label{ccl0}
\ee
while at late times they are
\be
x^{\prime i}_A = (\delta^i_j + \tfrac 12 h^{\infty\, i}_{\ \ \ j} ) X^j_A,\ \ \
x^{\prime i}_B = (\delta^i_j + \tfrac 12 h^{\infty\, i}_{\ \ \ j} ) X^j_B,
\label{ccl1}
\ee
as discussed in Sec.\ \ref{subsec:NewtArg} above.

To compute the generalized holonomy, it will be useful to recall results from
Sec.\ \ref{subsec:AffineTrans}.
First, recall that the generalized holonomy along a curve composed of
several segments is just the composition of the individual solutions to Eq.\
\eqref{eq:GenHolDef}.
Second, remember that the general solution can be written as the sum of a
homogeneous solution (i.e., the usual holonomy) and an inhomogeneous solution
(the part that is independent of the initial data), which allows the
two solutions to be computed independently.
Third, note that for geodesic curves the inhomogeneous part of the solution
is proportional to the tangent to the curve at the endpoint.
Thus, the generalized holonomy can be found by computing the affine transport
in four steps ($\mathcal P$ to $\mathcal R$ to $\mathcal S$ to $\mathcal Q$
to $\mathcal P$), while computing the inhomogeneous and homogeneous parts
separately.

\subsubsection{Calculation of the inhomogeneous solution}

\begin{itemize}

\item $\mathcal P$ to $\mathcal R$: We transport the initial vector
${\vec \xi}_{\mathcal P} = 0$ along the geodesic from $\mathcal P$ to
$\mathcal R$ using the affine transport law (\ref{eq:GenHolDef}) with
$\alpha =1$.  The result is
$$
{\vec \xi}_{\mathcal R} = (\delta t) \partial_T = (\delta t) \partial_t,
$$
where $\delta t$ is the interval of $A$'s
proper time between ${\cal P}$ and ${\cal R}$.
In the second equation, we have transformed from TT coordinates to the
inertial coordinates.

\item $\mathcal R$ to $\mathcal S$: Next, we use the vector
${\vec \xi}_{\mathcal R}$ as an initial condition for the affine transport
along the straight line extending from $\mathcal R$ to $\mathcal S$ in
the flat spacetime region after the burst.
It is easiest to perform this computation in the inertial coordinates
$(t,x^i)$.  The result is
$$
{\vec \xi}_{\cal S} = (\delta t) \partial_t + (x_B^{\prime i} -
x_A^{\prime i})
\frac{\partial}{\partial x^i}.
$$
Next we use Eqs.\ (\ref{ccl0}) and (\ref{ccl1}) and transform back to the TT coordinates,
giving
$$
{\vec \xi}_{\cal S} = (\delta t) \partial_T + (x_B^{i} -
x_A^{i})
\frac{\partial}{\partial X^i}.
$$
(Here in the spatial components, there was a cancellation between a
factor of ${\bf 1} + \tfrac 12 {\bf h}^\infty$ and its inverse.)

\item $\mathcal S$ to $\mathcal Q$: This part of the affine transport removes
the timelike component of the vector, and it transforms the spatial part
of $\xi^a_{\mathcal S}$ because the spatial vectors change under parallel
transport.
As a result, the outcome of the transport is
$$
{\vec \xi}_{\cal Q} = (\delta^i_j  + \tfrac 12 h^{\infty \, i}_{\ \ \
  j} ) (x^j_B - x^j_A) \frac{\partial}{\partial X^i}.
$$

\item $\mathcal Q$ to $\mathcal P$: The affine transport takes place along
a straight line in a flat spacetime region, and so
its net effect is to add the corresponding displacement vector along
the line.  The final result at ${\cal P}$ gives the inhomogeneous
piece of the general solution
\be
{\vec {\Delta \xi}} = {\vec \xi}_{\cal P}
= \tfrac 12 h^{\infty \, i}_{\ \ \
  j}  (x^j_B - x^j_A) \frac{\partial}{\partial X^i}.
\label{ans4a}
\ee

\end{itemize}

\subsubsection{Homogeneous solution and the generalized holonomy}

It is not too difficult to see that the holonomy of parallel transport
is the identity map,
\be
\Lambda^a_{\ b} = \delta^a_{\ b}
\label{ans4b}
\ee
for the curve shown in Fig.\ \ref{fig:Observers} even though the spacetime has
nontrivial curvature.
It follows from the fact that the parallel transport is trivial in the flat
regions of spacetime and that it is identical on the two worldlines of the
two different observers.
Consequently, the inhomogeneous solution is the only relevant part of the
generalized holonomy.

\subsubsection{Relation to the memory effect and the observer dependence
of angular momentum}

In this example, the generalized holonomy is directly related to the change
in proper distance between the two observers that arises from the solution
to the equation of geodesic deviation (the usual physical effect of the
gravitational-wave memory).
This leads to an observer dependence in angular momentum which is given by
$\delta J^{\alpha\beta} = 2 \Delta \xi^{[\alpha} P^{\beta]}$, from
Eqs.\ (\ref{eq:AngMomObsDep}) and (\ref{ans4b}).
Using the result (\ref{ans4a}) this corresponds to
an observer dependence of the spatial angular momentum
of the spatial angular
momentum of $-\tfrac{1}{2} \varepsilon_{ijk} h^{\infty \,j}_{\ \ \ \ l} \delta x^l p^k$,
where $p^k$ is the spatial momentum and $\delta {\bf x} = {\bf x}_B -
{\bf x}_A$.
This is precisely the result (\ref{eq:NewtJdiff})
found in Sec.\ \ref{subsec:NewtArg}.

\subsection{Generalized holonomy for a gravitational wave at large radius}
\label{subsec:GenHolGen}

For a gravitational wave propagating radially outward from a pointlike source,
the computation of the generalized holonomy is very similar to that of the
plane wave, but the expressions are somewhat lengthier.
The linearized metric of this spacetime has the same form as that of
Eq.\ (\ref{metric00}), but the function $h_{\alpha\beta}(t-z)$ gets
replaced by an outgoing wave solution in spherical polar coordinates.
The most common form of this metric is given in Lorentz gauge---see, e.g.,
Eqs.\ (8.13a)--(8.13c) of \cite{Thorne1980}---which is often expressed as a
sum of terms proportional to mass and current multipoles and the time
derivatives of the multipoles.
To compute the generalized holonomy, only the leading-order terms in a
series in $1/r$ will be needed.
In addition, it will be most useful to express the metric perturbation in a
TT gauge rather than Lorentz gauge.

\subsubsection{Transverse-traceless metric perturbation}

The quickest way to compute the TT metric perturbation is to compute the
Riemann tensor and use the fact that the TT metric perturbation is related
to the gauge-invariant Riemann tensor (at linear order in the metric
perturbation) via the relation
\begin{equation}
R_{0i0j} = \ddot h_{ij}^{\rm TT} \; ,
\label{eq:MetricRiemann}
\end{equation}
where the pair of dots over $h_{ij}^{\rm TT}$ indicates taking two time
derivatives.
The metric perturbation can be found by integrating Eq.\
\eqref{eq:MetricRiemann} twice with respect to time.
In coordinates $(u,r,\theta,\varphi)$ where $u = t - r$, and
starting from Eqs.\ (8.13a)--(8.13c) of \cite{Thorne1980},
the result is
\begin{align}
h_{ij}^{\rm TT} = & \frac{1}{r} \sum_{\ell=2}^\infty \bigg\{ \frac{1}{\ell !}
[4 n_{(i}\stackrel{(\ell)}{\mathcal I}_{j)A_{\ell-1}}n^{A_{\ell-1}}
- 2 \stackrel{(\ell)}{\mathcal I}_{ij A_{\ell-2}}
n^{A_{\ell-2}} \nonumber \\
&  - (\delta_{ij} + n_i n_j) \stackrel{(\ell)}{\mathcal I}_{A_\ell}n^{A_\ell}]
+ \frac{4\ell}{(\ell+1)!} n^q \times \nonumber \\
& [n_{(i}\varepsilon_{j)pq}\stackrel{(\ell)}{\mathcal S}_{p A_{\ell-1}}
n^{A_{\ell-1}} - \varepsilon_{pq(i}
\stackrel{(\ell)}{\mathcal S}_{j)p A_{\ell-2}} n^{A_{\ell-2}}] \bigg\}
\nonumber \\
& + O(1/r^2)\; .
\label{eq:MetricTT}
\end{align}
Here $\mathcal I_{A_\ell} = \mathcal I_{A_\ell}(u)$ is an $\ell$-pole mass moment and
$\mathcal S_{A_\ell} = \mathcal S_{A_\ell}(u)$ is an $\ell$-pole current moment, which are symmetric
trace-free (STF) tensors with $\ell$ indices (the subscript $A_\ell$ is one
notation used to represent $\ell$ spatial indices).
The notation $(\ell)$ above the symbols for the moments means the
$\ell^{\rm th}$ derivative with respect to $u$.
The vector $n^i$ is a unit radial vector (i.e., $x^i/r$) and $n^{A_\ell}$ is
the tensor product of $\ell$ radial unit vectors.

\subsubsection{Multipoles and coordinate change after the burst}

As in the example of the plane wave, it will be assumed that before a retarded
time $u\equiv t-r=0$ all the multipoles vanish; they are dynamical between $0$
and $u_f$; and after the retarded time $u_f$, the spacetime is again Minkowski,
but some of the multipoles and their time derivatives can have nonzero
constant values which correspond to the gravitational-wave memory.
Interestingly, only certain multipoles can go to constant values and still
have the spacetime be Minkowski (and, hence, stationary).
Specifically, the $\ell^{\rm th}$ time derivative of the mass-multipole STF
tensors $\mathcal I_{A_\ell}$ can take nonzero values
whereas the equivalent time derivatives of the current multipoles
$\mathcal S_{A_\ell}$ cannot asymptote to a nonzero value
and still be Minkowski space.
This seems to be closely related to the fact that there is no magnetic-type
memory from physically realistic sources \cite{Winicour2014}.

First, consider just the mass multipoles, and assume that the $\ell^{\rm th}$
time derivatives go to constant values.
A short calculation can show that the generator of linearized gauge
transformations below can remove the constant time derivatives of the mass
multipoles after the burst of waves:
\begin{subequations}
\label{eq:MassMomentGauge}
\begin{align}
\Xi_0 = & \sum_{\ell=2}^\infty \frac{\ell+2}{\ell (\ell !)}
\stackrel{(\ell)}{\mathcal I}_{A_\ell}n^{A_\ell} \; , \\
\Xi_i = & - \sum_{\ell=2}^\infty \frac{1}{\ell !} \bigg[ \frac{1}{\ell-1}
\stackrel{(\ell)}{\mathcal I}_{iA_{\ell-1}}n^{A_{\ell-1}} + \frac{1}{2}
\stackrel{(\ell)}{\mathcal I}_{A_\ell}n^{A_\ell} n_i\nonumber \\
& -\frac{(\ell+2)}{2r}( \stackrel{(\ell-1)}{\mathcal I}_{iA_{\ell-1}}
n^{A_{\ell-1}} - \stackrel{(\ell-1)}{\mathcal I}_{A_\ell}n^{A_\ell}n_i )
\bigg] \; .
\end{align}
\end{subequations}
For the current multipoles, the only linearized gauge generator that can be
constructed from the $\ell^{\rm th}$ time derivative of $\mathcal S_{A_\ell}$,
the radial vectors $n^i$, and the antisymmetric tensor $\varepsilon_{ipq}$
would be proportional to the following:
\begin{equation}
\Xi^{(\mathcal S)}_i = \sum_{\ell=2}^\infty \varepsilon_{ipq}
\stackrel{(\ell)}{\mathcal S}_{p A_{\ell-1}} n^{A_{\ell-1}} n^q \; .
\end{equation}
A second quick calculation will show that this transformation does {\em not}
make the spacetime flat.
As a result, we will require that the multipoles satisfy the conditions
\begin{equation}
\stackrel{(\ell)}{\mathcal I}_{A_\ell} = {\rm const.} \quad {\rm and}
\quad \stackrel{(\ell)}{\mathcal S}_{A_\ell} = 0 \; ,
\label{eq:MultipoleCond}
\end{equation}
when $u>u_f$.

With the metric determined by Eq.\ \eqref{eq:MetricTT}, subject to the
condition \eqref{eq:MultipoleCond}, the gauge transformation
\eqref{eq:MassMomentGauge} is sufficient to define Minkowski coordinates
after the pulse of waves via the relation
\be
y^\alpha = x^\alpha + \Xi^\alpha.
\label{flatc}
\ee
This also provides the necessary information to compute the generalized
holonomy.
As in the previous plane-wave example, we will split the calculation into
the inhomogeneous and homogeneous parts, which are treated in the next
subparts, respectively.

\subsubsection{Calculation of the inhomogeneous solution}

\begin{itemize}

\item $\mathcal P$ to $\mathcal R$: This is identical to the equivalent
calculation involving the gravitational plane wave: the vector after affine
transport is $\xi^\alpha_{\mathcal R} = (\delta t) u^\alpha$, where
${\vec u} = \partial_t$.
As before, it is helpful to transform to the flat coordinates
$y^\alpha$ after the pulse, defined by Eqs.\ (\ref{flatc}) and (\ref{eq:MassMomentGauge}).
This introduces two new terms into the result:
$\xi^{\alpha'}_{\mathcal R} = (\delta t-\Xi_{\mathcal R}^{0'})u^{\alpha'} -
\Xi_{\mathcal R}^{\alpha'}{}_{,0'}\delta t$.
Here we use primes to denote tensor components in the Minkowski
coordinates, $x^{\alpha'} = y^\alpha$.

\item $\mathcal R$ to $\mathcal S$: In the flat Minkowski space after the
burst, the affine transport gives $\xi^{\alpha'}_{\mathcal S} =
(\delta t-\Xi_{\mathcal R}^{0'})u^{\alpha'}-\Xi_{\mathcal R}^{\alpha'}{}_{,0'}
\delta t + \delta^{\alpha'}{}_{i'} \delta y^{i'}$.
Changing back to the $x^\alpha$ coordinates alters the spatial part of the
vector so that
$\xi^\alpha_{\mathcal S} = u^\alpha(\delta t+\delta\Xi^0) + \delta
\Xi^\alpha{}_{,0}\delta t + \Xi_{\mathcal S}^{\alpha}{}_{,i}\delta x^i + \delta^\alpha{}_i(\delta x^i + \delta \Xi^i)$.
Here $\delta \Xi^\alpha = \Xi^\alpha_{\mathcal S} - \Xi^\alpha_{\mathcal R}$
has been defined.

\item $\mathcal S$ to $\mathcal Q$: Transporting back through the burst
changes the spatial part of the vector to $\xi^\alpha_{\mathcal Q} =
\delta\Xi^0 u^\alpha +\delta \Xi^\alpha{}_{,0}\delta t + \tfrac{1}{2}
(\Xi_{\mathcal S}^{\alpha}{}_{,i}-\Xi^{\mathcal S}_i{}^{,\alpha})
\delta x^i + \delta^\alpha{}_i (\delta x^i + \delta \Xi^i)$, where the change
occurred from the parallel transport of the affine frame back to the original
point and where the fact that $h^\alpha{}_i = \Xi^{\alpha}{}_{,i} +
\Xi_i{}^{,\alpha}$ was used to simplify the change in the spatial part of
the vector.

\item $\mathcal Q$ to $\mathcal P$: Along this flat geodesic in the Minkowski
space prior to the burst, the affine transport adds the displacement vector
$\delta x^i$ to the result of $\xi^\alpha_{\mathcal Q}$.
Thus, the complete inhomogeneous solution is
\be
\Delta \xi^\alpha = \delta \Xi^0 u^\alpha +\delta
\Xi^\alpha{}_{,0}\delta t + \tfrac{1}{2}(\Xi_{\mathcal S}^{\alpha}{}_{,i} -
\Xi^{\mathcal S}_i{}^{,\alpha}) \delta x^i + \delta^\alpha{}_i  \delta \Xi^i.
\ee
\end{itemize}

We will discuss the relationship between the terms that appear in
$\Delta \xi^a_{\mathcal P}$ and the gravitational-wave memory in more detail
below.

\subsubsection{Homogeneous solution and the generalized holonomy}

The calculation of the homogeneous part of the solution is simpler than that
of the inhomogeneous portion above.
The first set of nontrivial terms comes from the parallel transport along
the worldline extending from $\mathcal P$ to $\mathcal R$, and from the
coordinate change at $\mathcal R$.
For an arbitrary initial condition $\xi^\alpha_{(0)}$, this vector will be
modified by an amount $-\tfrac{1}{2} (\Xi_{\mathcal R}^{\alpha}{}_{,\beta} -
\Xi^{\mathcal R}_\beta{}^{,\alpha})\xi^\beta_{(0)}$.
There will be a similar contribution with the opposite sign involving
quantities at the point $\mathcal S$ from the parallel transport along the
worldline from $\mathcal S$ to $\mathcal Q$ and the coordinate change at
$\mathcal S$.
Thus, the part of the holonomy that differs from the identity is given by
$\tfrac{1}{2} (\Xi^{\alpha}{}_{,\beta}-\Xi_\beta{}^{,\alpha})_{\mathcal{SR}}
\xi^\beta_{(0)}$, where the subscript $\mathcal{SR}$ implies it is the
difference of the values at the quantities at the coordinate points
$\mathcal S$ and $\mathcal R$, transported back to $\mathcal P$.

From the expression for $\Xi^\alpha$ in the gauge transformation
\eqref{eq:MassMomentGauge}, it is possible to show that the generalized
holonomy has a homogeneous piece in the form of a local infinitesimal Lorentz
transformation that scales as $1/r$, and an inhomogeneous part that contains
terms independent of $\delta x$ and $\delta t$ that are zeroth order in $1/r$
(and also terms that go as $1/r$, which we will not show), in addition to
terms that scale as $\delta x/r$, and $\delta t/r$.
For the inhomogeneous part, these terms will be labeled by
$\Delta\xi^\alpha_{(1)}$, $\Delta \xi^\alpha_{(\delta x/r)}$, and
$\Delta \xi^\alpha_{(\delta t/r)}$.
The zeroth-order terms come from $-\delta \Xi^\alpha$, whereas the
terms of order $\delta t/r$ and $\delta x/r$ come from the terms $\delta
\Xi^\alpha{}_{,0}\delta t$ and $\tfrac{1}{2} (\Xi_{\mathcal S}^{\alpha}{}_{,i}
-\Xi^{\mathcal S}_i{}^{,\alpha})\delta x^i$, respectively.
These terms are
\begin{subequations}
\label{eq:MassMomentTrans}
\begin{align}
\Delta \xi^0_{(1)} = & \sum_{\ell=2}^\infty \frac{\ell+2}{\ell(\ell !)}
(\stackrel{(\ell)}{\mathcal I}_{A_\ell}n^{A_\ell})_{\mathcal{SR}} \; , \\
\Delta \xi_i^{(1)} = & -\sum_{\ell=2}^\infty \frac{1}{\ell !} \bigg[
\frac{1}{\ell-1} (\stackrel{(\ell)}{\mathcal I}_{iA_{\ell-1}}
n^{A_{\ell-1}})_{\mathcal{SR}} \nonumber \\
& + \frac{1}{2} (\stackrel{(\ell)}{\mathcal I}_{A_\ell}n^{A_\ell}
n_i)_{\mathcal{SR}} \bigg] \; , \\
\Delta \xi_i^{(\delta t/r)} = & -\delta t\sum_{\ell=2}^\infty
\frac{\ell+2}{2(\ell!)} [(\stackrel{(\ell)}{\mathcal I}_{iA_{\ell-1}}
n^{A_{\ell-1}}/r)_{\mathcal{SR}} \nonumber \\
&-(\stackrel{(\ell)}{\mathcal I}_{A_\ell} n^{A_\ell} n_i/r)_{\mathcal{SR}}]
\; ,\\
\Delta \xi^0_{(\delta x/r)} = & -\frac{\delta x^i}{r_{\mathcal S}}
\sum_{\ell=2}^\infty \frac{\ell+2}{\ell!}
[(\stackrel{(\ell)}{\mathcal I}_{iA_{\ell-1}}n^{A_{\ell-1}})_{\mathcal S}
\nonumber \\
& - (\stackrel{(\ell)}{\mathcal I}_{A_\ell} n^{A_\ell} n_i)_{\mathcal S}]
 \; ,\\
\Delta \xi_i^{(\delta x/r)} = & \frac{2\delta x^j}{r_{\mathcal S}}
\sum_{\ell=2}^\infty \frac{1}{\ell!} (n_{[i}
\stackrel{(\ell)}{\mathcal I}_{j]A_{\ell-1}} n^{A_{\ell-1}})_{\mathcal S} \; .
\end{align}
\end{subequations}
In the expression above, the subscript $\mathcal{SR}$ means to take the
difference of the quantity within parentheses evaluated at the values of the
coordinate points $\mathcal S$ and $\mathcal R$.

The local infinitesimal Lorentz transformation, which will be denoted as
$\omega_{\alpha\beta} = \omega_{[\alpha\beta]}$ is strictly of order $1/r$
and can be written as
\begin{subequations}
\begin{align}
\omega_{i0} = & \sum_{\ell=2}^\infty \frac{\ell+2}{\ell!}
[(\stackrel{(\ell)}{\mathcal I}_{iA_{\ell-1}}n^{A_{\ell-1}}/r)_{\mathcal{SR}}
- (\stackrel{(\ell)}{\mathcal I}_{A_\ell} n^{A_\ell} n_i/r)_{\mathcal{SR}}]
\nonumber \\
 & + O(1/r^2) \; ,\\
\omega_{ij} = & \sum_{\ell=2}^\infty \frac{2}{\ell!}
(n_{[i} \stackrel{(\ell)}{\mathcal I}_{j]A_{\ell-1}}
n^{A_{\ell-1}}/r)_{\mathcal{SR}} + O(1/r^2) \; .
\end{align}
\end{subequations}

\subsubsection{Relation to the memory effect and the observer dependence of
angular momentum}

The relation between the generalized holonomy and the physical effects
associated with the gravitational-wave memory is somewhat more involved than
it was for a gravitational plane wave.
The term $\delta \Xi^i$ is a measure of the change in distance between
the observers that occurs from the memory.
In addition, the part $\delta \Xi^0$ gives information about the difference
in proper time measured by the two observers that is a result of the memory
of the gravitational-wave burst.
The other term $\tfrac{1}{2} (\Xi_{\mathcal S}^{\alpha}{}_{,i} -
\Xi^{\mathcal S}_i{}^{,\alpha})\delta x^i$ takes into account a boosting and
rotation of the spatial displacement vector along the other observer's
worldline from the wave's memory, and the part $\delta \Xi^\alpha{}_{,0}
\delta t$ represents a relative change in the tangent to the observers'
worldlines from the memory.

Because the inhomogeneous solution has a zeroth-order piece in $1/r$, the
center of mass and the angular momentum will have an observer dependence
with a magnitude of order $P^0 \Delta\xi^i_{(1)} + P^i \Delta\xi^0_{(1)}$ and
$\epsilon_{ijk} \Delta\xi^j_{(1)} P^k$, respectively, where $P^a$ is the
4-momentum of the source.
For separations for which $\delta x$ is of order $r$, then the terms
$\Delta \xi^a_{(\delta x/r)}$ will also have leading-order contributions to
the observer dependence of the center of mass and of the angular momentum
of the form $P^0 \Delta\xi^i_{(\delta x/r)} + P^i \Delta\xi^0_{(\delta x/r)}$
and $\epsilon_{ijk} \Delta\xi^j_{(\delta x/r)} P^k$, respectively.
Similarly, for times $\delta t$ of order the light-travel time to the source
(i.e., of order $r$), then there will be additional observer dependence
from terms of the form $P^0 \Delta\xi^i_{(\delta t/r)}$
and $\epsilon_{ijk} \Delta\xi^j_{(\delta t/r)} P^k$.

Equations \eqref{eq:ydef} and \eqref{eq:Jabdef}
imply that the angular momentum tensor $J^{ab}$ will have terms proportional
to $r$ at large radii: specifically, it is the orbital-like part of the
angular momentum $2 y^{[a} P^{b]} = -2r n^{[a} P^{b]}$ that has this scaling.
When the angular momentum transforms by Eq.\ \eqref{eq:AngMomGenHol},
the $1/r$ parts of the holonomy will induce a change in the
angular momentum that is of order unity in a series in $1/r$.
These terms will have the form
$\delta J^{ab} = 2 (\omega^a{}_c y^{[c} P^{b]} - \omega^a{}_c y^{[b}
P^{c]})$.
The lowest-order part of the 4-momentum will still be unambiguous, and
any observer dependence will be a relative $1/r$ effect.

\section{Conclusions}
\label{sec:Conclusions}

In this paper, we noted that bursts of gravitational waves cause
spatially separated observers to disagree on their changes in
displacement and therefore to disagree on their measured special-relativistic
angular momenta of a source.
This observer dependence of angular momentum is related to the
gravitational-wave memory of the pulse of waves.
We derived this phenomenon first in a simple context of
linearized plane waves, and later in a more systematic and covariant framework.

We defined a procedure by which observers could measure a type
of special-relativistic linear and angular momentum at their
locations, from the spacetime geometry in their vicinity.
The procedure gives the correct result when the spacetime is linear and
stationary, and the measurement takes place near future null infinity.
We estimated the errors in the procedure when the spacetime is
nonlinear, dynamical, or the source is not isolated.

To compare angular momentum at different spacetime points, we defined
a transport equation, the {\it affine transport}, which is a slight
generalization of parallel transport.
The transport around a closed curve, the {\it generalized holonomy},
consists of a Poincar\'e transformation, rather than a Lorentz
transformation as for a normal holonomy.  The generalized holonomy
contains an inhomogeneous displacement term.
The extent to which the generalized holonomy is nontrivial is a
measure of how much spacetime curvature prevents different observers from arriving
at a consistent definition of linear and angular momentum.

For two freely falling observers, who encounter a burst of
gravitational waves, we showed that there are four independent
observables that can be nontrivial when the burst has departed
and that can be considered to be types of ``gravitational-wave memory.''
There is the usual displacement memory, a residual relative boost,
a relative rotation, and a difference in elapsed proper time between
the two observers.  These four observables are all encoded in the
generalized holonomy around a suitably defined closed loop in
spacetime.  Thus, we clarified and generalized the often-noted
close relation between gravitational-wave memory and observer dependence of angular momentum.

Finally, we performed explicit computations in two different specific
contexts that illustrate
the relationships between generalized
holonomy, observer dependence of angular momentum, and
gravitational-wave memory.
The first context was a plane gravitational wave with memory passing through 
flat spacetime, and the second was an outgoing linearized gravitational wave 
near future null infinity.
The plane wave only showed the displacement memory effect, but the
multipolar gravitational wave displayed all four of the physical observables
associated with the memory.

Although our goal was to provide physical insight into the nature of the BMS 
group, the generalized holonomy tool does not quite achieve this goal:
in Appendix \ref{app:GenHolSchw}, we show that the generalized holonomy can be 
nontrivial for certain spacelike curves in the Schwarzschild spacetime, even 
as the curves tend to spatial infinity.
Hence, observers along this curve would find their measured angular momentum
to be observer dependent, even though angular momentum is well defined
in Schwarzschild (as the BMS group has a preferred Poincar\'e subgroup
in stationary spacetimes).
Thus, our prescription for assessing observer dependence
in angular momentum not only captures BMS/memory ambiguities in angular
momentum but also reflects other, more trivial effects of spacetime
curvature on angular momentum measurements.
Finding a method to isolate just the BMS ambiguities is a topic we will
investigate in future work.

Because the affine transport law defines a way to compare other vectors in
addition to the angular momentum at different spacetime points, it could find
application to other problems.
For example, if a burst of gravitational waves passes through a
post-Newtonian spacetime, the momenta and angular momenta of the particles
that enter into the post-Newtonian equations of motion could differ before
and after the burst.
The affine transport may be useful for deriving a prescription
for matching the post-Newtonian spacetimes before and after the bursts,
in a manner that allows one to compute the motion of an $N$-body system.

\acknowledgments

We thank Justin Vines for reading an earlier draft of this paper and for
providing helpful comments and also Abhay Ashtekar, Mike Boyle, and Leo Stein
for helpful discussions.
This work was supported in part by NSF grants No.\ PHY-1404105 and 
No.\ PHY-1068541.

\appendix
\section{Angular Momentum Transport Laws}
\label{sec:alternative}

In the body of this paper, we introduced a method of transporting a
pair of tensors $(P^a,J^{ab})$ along a curve from one point to another
in a curved spacetime.  In this Appendix, we show that the transport
method is equivalent to solving the following simple set of
differential equations along the curve:
\be
k^a \nabla_a P^b = 0, \ \
k^a \nabla_a J^{bc} = 2 P^{[b} k^{c]}.
\label{eq:transport}
\ee
The equivalence between the two methods was pointed out to us by
Justin Vines \cite{Vines2014b}.

Suppose we have a curve $x^\alpha = x^\alpha(\lambda)$ that joins at
point ${\cal P}$ at $\lambda =0$ to another point ${\cal Q}$ at
$\lambda = 1$.  We introduce an orthonormal basis of vectors ${\vec
  e}_{\hat \alpha}$ at ${\cal Q}$, and extend it along the curve by
parallel transport.  We decompose the 4-momentum and angular
momentum on this basis as
\be
P^a = P^{{\hat \alpha}} e_{\hat \alpha}^a, \ \ \ \ \
J^{ab} = J^{{\hat \alpha}{\hat \beta}} e_{\hat \alpha}^a e_{\hat \beta}^b.
\ee
The transport equations (\ref{eq:transport}), when written in terms of
this basis, become
\begin{subequations}
\begin{eqnarray}
\frac{d}{d\lambda} P^{\hat \alpha} &=&0, \\
\frac{d}{d\lambda} J^{{\hat \alpha}{\hat \beta}} &=& P^{\hat \alpha}
k^{\hat \beta} - P^{\hat \beta} k^{\hat \alpha}.
\label{eq:sss}
\end{eqnarray}
\end{subequations}
The first of these gives $P^{\hat \alpha}(\lambda) = P^{\hat \alpha}_0
= $ constant.  We now make the ansatz for the angular momentum
solution
\be
J^{{\hat \alpha}{\hat \beta}}(\lambda) = J_0^{{\hat \alpha}{\hat \beta}} 
+ P_0^{{\hat \alpha}} \chi^{\hat \beta}(\lambda) -
 P_0^{{\hat \beta}} \chi^{\hat \alpha}(\lambda),
\label{eq:ansatz}
\ee
for some vector $\chi^{\hat \alpha}(\lambda)$, where
$J_0^{{\hat \alpha}{\hat \beta}}$ is the initial value of 
$J^{{\hat \alpha}{\hat \beta}}$ at $\lambda = 0$.
Using this ansatz we see that the differential equation (\ref{eq:sss})
will be satisfied if $\chi^{\hat \alpha}$ vanishes at ${\cal P}$ and
satisfies
\be
\frac{d}{d\lambda} \chi^{\hat \alpha} = k^{\hat \alpha}.
\ee
This differential equation coincides with the differential equation
(\ref{eq:GenHolDef}) that defines the generalized parallel transport,
for the case $\alpha = +1$.
By comparing with Eq.\ (\ref{eq:ATdef}) we find
\be
\chi^{\hat \alpha}({\cal Q}) = \Delta \xi^{\hat \alpha}_{\mathcal{PQ}}.
\ee
Substituting this result into the ansatz (\ref{eq:ansatz}) gives
an expression for the angular momentum at ${\cal Q}$ which agrees with
Eq.\ (\ref{eq:AngMomObsDep}), establishing the result.

\section{Generalized holonomy of a spacelike curve in the Schwarzschild
spacetime}
\label{app:GenHolSchw}

\begin{figure}
\includegraphics[width=\columnwidth]{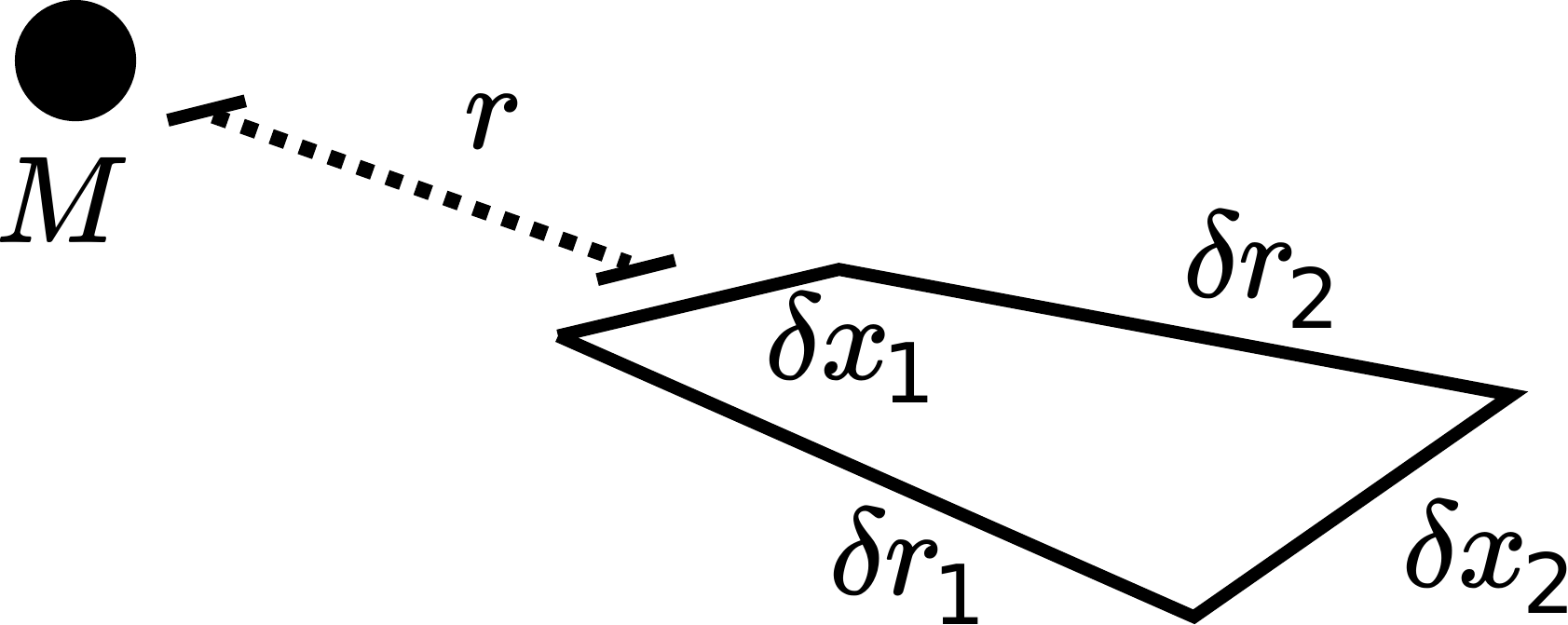}
\caption{Curve used to compute the generalized holonomy in the Schwarzschild
spacetime of mass $M$ at large radii.
The lengths of the four segments that compose the curve, $\delta r_1$,
$\delta r_2$, $\delta x_1$, and $\delta x_2$ are all of order $r$, where
$r\gg M$ is the closest distance to the source along the segment labeled
by $\delta x_1$.
This curve has a nontrivial generalized holonomy as $r$ goes to spatial
infinity, even though the Schwarzschild spacetime has a  well-defined
angular momentum.}
\label{fig:SchwCurve}
\end{figure}

In this Appendix, we compute the generalized holonomy of certain curves 
in the Schwarzschild spacetime and show that the generalized holonomy
{\it does not} become the identity in the limit when the curves asymptote to
future null infinity.  As discussed in the body of the paper, this property
implies that we cannot use the generalized holonomy as a tool to diagnose 
whether a given spacetime admits a well-defined angular momentum (since in 
stationary spacetimes the BMS group has a preferred Poincar\'e subgroup and 
so normal angular momentum is well defined).

The specific curve we consider is shown in Fig.\ \ref{fig:SchwCurve}.
Letting $M$ denote the mass of the spacetime and $r\gg M$ be the distance
to the closest point along the segment labeled by $\delta x_1$, we will
assume that the lengths of the four sides of the curve, $\delta x_1$,
$\delta r_1$, $\delta x_2$, and $\delta r_2$ are all of order $r$ (or
equivalently, the area enclosed by the curve is of order $r^2$).
The curvature at the loop will scale as $M/r^3$.
Because the holonomy associated with parallel transport scales as the
curvature times the area, when the vector transported has magnitude of order
$r$ (such as the displacement vector $y^a$), the vector will undergo
changes of order $M$.
Similarly, because the inhomogeneous part of the generalized holonomy
scales as curvature times the area to the three-halves power, the vector
$\Delta \xi^a$ will also be of order $M$.
Even as $r$ approaches spatial infinity, this estimate suggests that there
will be nontrivial generalized holonomy and observer dependence in angular
momentum.

We did, in fact, compute the exact generalized holonomy in the case in
which $\delta r_1$ and $\delta r_2$ are radial curves, and $\delta x_1$
and $\delta x_2$ are two coordinate lines between the endpoints of these
two curves, respectively.
The precise answer, while not particularly insightful, does indeed scale
as $M$ as $r$ goes to spatial infinity.
We conclude that the generalized holonomy is not
specifically linked to BMS ambiguities in angular momentum, and is more
generally a diagnostic of when spacetime curvature prevents observers
from consistently measuring and comparing a type of special-relativistic
angular momentum.

\bibliography{Refs}

\begin{thebibliography}{28}%
\makeatletter
\providecommand \@ifxundefined [1]{%
 \@ifx{#1\undefined}
}%
\providecommand \@ifnum [1]{%
 \ifnum #1\expandafter \@firstoftwo
 \else \expandafter \@secondoftwo
 \fi
}%
\providecommand \@ifx [1]{%
 \ifx #1\expandafter \@firstoftwo
 \else \expandafter \@secondoftwo
 \fi
}%
\providecommand \natexlab [1]{#1}%
\providecommand \enquote  [1]{``#1''}%
\providecommand \bibnamefont  [1]{#1}%
\providecommand \bibfnamefont [1]{#1}%
\providecommand \citenamefont [1]{#1}%
\providecommand \href@noop [0]{\@secondoftwo}%
\providecommand \href [0]{\begingroup \@sanitize@url \@href}%
\providecommand \@href[1]{\@@startlink{#1}\@@href}%
\providecommand \@@href[1]{\endgroup#1\@@endlink}%
\providecommand \@sanitize@url [0]{\catcode `\\12\catcode `\$12\catcode
  `\&12\catcode `\#12\catcode `\^12\catcode `\_12\catcode `\%12\relax}%
\providecommand \@@startlink[1]{}%
\providecommand \@@endlink[0]{}%
\providecommand \url  [0]{\begingroup\@sanitize@url \@url }%
\providecommand \@url [1]{\endgroup\@href {#1}{\urlprefix }}%
\providecommand \urlprefix  [0]{URL }%
\providecommand \Eprint [0]{\href }%
\providecommand \doibase [0]{http://dx.doi.org/}%
\providecommand \selectlanguage [0]{\@gobble}%
\providecommand \bibinfo  [0]{\@secondoftwo}%
\providecommand \bibfield  [0]{\@secondoftwo}%
\providecommand \translation [1]{[#1]}%
\providecommand \BibitemOpen [0]{}%
\providecommand \bibitemStop [0]{}%
\providecommand \bibitemNoStop [0]{.\EOS\space}%
\providecommand \EOS [0]{\spacefactor3000\relax}%
\providecommand \BibitemShut  [1]{\csname bibitem#1\endcsname}%
\let\auto@bib@innerbib\@empty
\bibitem [{\citenamefont {{Bondi}}\ \emph {et~al.}(1962)\citenamefont
  {{Bondi}}, \citenamefont {{van der Burg}},\ and\ \citenamefont
  {{Metzner}}}]{Bondi1962}%
  \BibitemOpen
  \bibfield  {author} {\bibinfo {author} {\bibfnamefont {H.}~\bibnamefont
  {{Bondi}}}, \bibinfo {author} {\bibfnamefont {M.~G.~J.}\ \bibnamefont {{van
  der Burg}}}, \ and\ \bibinfo {author} {\bibfnamefont {A.~W.~K.}\ \bibnamefont
  {{Metzner}}},\ }\href {\doibase 10.1098/rspa.1962.0161} {\bibfield  {journal}
  {\bibinfo  {journal} {Proc. R. Soc. Lond. A}\ }\textbf {\bibinfo {volume}
  {269}},\ \bibinfo {pages} {21} (\bibinfo {year} {1962})}\BibitemShut
  {NoStop}%
\bibitem [{\citenamefont {{Sachs}}(1962)}]{Sachs1962}%
  \BibitemOpen
  \bibfield  {author} {\bibinfo {author} {\bibfnamefont {R.~K.}\ \bibnamefont
  {{Sachs}}},\ }\href {\doibase 10.1098/rspa.1962.0206} {\bibfield  {journal}
  {\bibinfo  {journal} {Proc. R. Soc. Lond. A}\ }\textbf {\bibinfo {volume}
  {270}},\ \bibinfo {pages} {103} (\bibinfo {year} {1962})}\BibitemShut
  {NoStop}%
\bibitem [{\citenamefont {Sachs}(1962)}]{Sachs1962b}%
  \BibitemOpen
  \bibfield  {author} {\bibinfo {author} {\bibfnamefont {R.}~\bibnamefont
  {Sachs}},\ }\href {\doibase 10.1103/PhysRev.128.2851} {\bibfield  {journal}
  {\bibinfo  {journal} {Phys. Rev.}\ }\textbf {\bibinfo {volume} {128}},\
  \bibinfo {pages} {2851} (\bibinfo {year} {1962})}\BibitemShut {NoStop}%
\bibitem [{\citenamefont {Stewart}(1993)}]{Stewart1993}%
  \BibitemOpen
  \bibfield  {author} {\bibinfo {author} {\bibfnamefont {J.}~\bibnamefont
  {Stewart}},\ }\href@noop {} {\emph {\bibinfo {title} {Advanced General
  Relativity}}}\ (\bibinfo  {publisher} {Cambridge University Press},\ \bibinfo
  {address} {Cambridge, England},\ \bibinfo {year} {1993})\BibitemShut
  {NoStop}%
\bibitem [{\citenamefont {{Bondi}}(1960)}]{Bondi1960}%
  \BibitemOpen
  \bibfield  {author} {\bibinfo {author} {\bibfnamefont {H.}~\bibnamefont
  {{Bondi}}},\ }\href {\doibase 10.1038/186535a0} {\bibfield  {journal}
  {\bibinfo  {journal} {Nature}\ }\textbf {\bibinfo {volume} {186}},\ \bibinfo
  {pages} {535} (\bibinfo {year} {1960})}\BibitemShut {NoStop}%
\bibitem [{\citenamefont {{Ashtekar}}\ and\ \citenamefont
  {{Streubel}}(1979)}]{Ashtekar1979}%
  \BibitemOpen
  \bibfield  {author} {\bibinfo {author} {\bibfnamefont {A.}~\bibnamefont
  {{Ashtekar}}}\ and\ \bibinfo {author} {\bibfnamefont {M.}~\bibnamefont
  {{Streubel}}},\ }\href {\doibase 10.1063/1.524242} {\bibfield  {journal}
  {\bibinfo  {journal} {J. Math. Phys.}\ }\textbf {\bibinfo {volume} {20}},\
  \bibinfo {pages} {1362} (\bibinfo {year} {1979})}\BibitemShut {NoStop}%
\bibitem [{\citenamefont {{Dray}}\ and\ \citenamefont
  {{Streubel}}(1984)}]{Dray1984}%
  \BibitemOpen
  \bibfield  {author} {\bibinfo {author} {\bibfnamefont {T.}~\bibnamefont
  {{Dray}}}\ and\ \bibinfo {author} {\bibfnamefont {M.}~\bibnamefont
  {{Streubel}}},\ }\href {\doibase 10.1088/0264-9381/1/1/005} {\bibfield
  {journal} {\bibinfo  {journal} {Classical Quantum Gravity}\ }\textbf
  {\bibinfo {volume} {1}},\ \bibinfo {pages} {15} (\bibinfo {year}
  {1984})}\BibitemShut {NoStop}%
\bibitem [{\citenamefont {Wald}\ and\ \citenamefont {Zoupas}(2000)}]{Wald2000}%
  \BibitemOpen
  \bibfield  {author} {\bibinfo {author} {\bibfnamefont {R.~M.}\ \bibnamefont
  {Wald}}\ and\ \bibinfo {author} {\bibfnamefont {A.}~\bibnamefont {Zoupas}},\
  }\href {\doibase 10.1103/PhysRevD.61.084027} {\bibfield  {journal} {\bibinfo
  {journal} {Phys. Rev. D}\ }\textbf {\bibinfo {volume} {61}},\ \bibinfo
  {pages} {084027} (\bibinfo {year} {2000})}\BibitemShut {NoStop}%
\bibitem [{\citenamefont {Adamo}\ \emph {et~al.}(2012)\citenamefont {Adamo},
  \citenamefont {Newman},\ and\ \citenamefont {Kozameh}}]{Adamo2012}%
  \BibitemOpen
  \bibfield  {author} {\bibinfo {author} {\bibfnamefont {T.~M.}\ \bibnamefont
  {Adamo}}, \bibinfo {author} {\bibfnamefont {E.~T.}\ \bibnamefont {Newman}}, \
  and\ \bibinfo {author} {\bibfnamefont {C.}~\bibnamefont {Kozameh}},\ }\href
  {\doibase 10.12942/lrr-2012-1} {\bibfield  {journal} {\bibinfo  {journal}
  {Living Rev. Relativity}\ }\textbf {\bibinfo {volume} {15}},\ \bibinfo
  {pages} {1} (\bibinfo {year} {2012})}\BibitemShut {NoStop}%
\bibitem [{\citenamefont {{Moreschi}}(1986)}]{Moreschi1986}%
  \BibitemOpen
  \bibfield  {author} {\bibinfo {author} {\bibfnamefont {O.~M.}\ \bibnamefont
  {{Moreschi}}},\ }\href {\doibase 10.1088/0264-9381/3/4/006} {\bibfield
  {journal} {\bibinfo  {journal} {Classical Quantum Gravity}\ }\textbf
  {\bibinfo {volume} {3}},\ \bibinfo {pages} {503} (\bibinfo {year}
  {1986})}\BibitemShut {NoStop}%
\bibitem [{\citenamefont {{Moreschi}}(1988)}]{Moreschi1988}%
  \BibitemOpen
  \bibfield  {author} {\bibinfo {author} {\bibfnamefont {O.~M.}\ \bibnamefont
  {{Moreschi}}},\ }\href {\doibase 10.1088/0264-9381/5/3/004} {\bibfield
  {journal} {\bibinfo  {journal} {Classical Quantum Gravity}\ }\textbf
  {\bibinfo {volume} {5}},\ \bibinfo {pages} {423} (\bibinfo {year}
  {1988})}\BibitemShut {NoStop}%
\bibitem [{\citenamefont {{Rizzi}}(1998)}]{1998PhRvL..81.1150R}%
  \BibitemOpen
  \bibfield  {author} {\bibinfo {author} {\bibfnamefont {A.}~\bibnamefont
  {{Rizzi}}},\ }\href {\doibase 10.1103/PhysRevLett.81.1150} {\bibfield
  {journal} {\bibinfo  {journal} {Phys. Rev. Lett.}\ }\textbf {\bibinfo
  {volume} {81}},\ \bibinfo {pages} {1150} (\bibinfo {year}
  {1998})}\BibitemShut {NoStop}%
\bibitem [{\citenamefont {{Dain}}\ and\ \citenamefont
  {{Moreschi}}(2000)}]{Dain2000}%
  \BibitemOpen
  \bibfield  {author} {\bibinfo {author} {\bibfnamefont {S.}~\bibnamefont
  {{Dain}}}\ and\ \bibinfo {author} {\bibfnamefont {O.~M.}\ \bibnamefont
  {{Moreschi}}},\ }\href {\doibase 10.1088/0264-9381/17/18/305} {\bibfield
  {journal} {\bibinfo  {journal} {Classical Quantum Gravity}\ }\textbf
  {\bibinfo {volume} {17}},\ \bibinfo {pages} {3663} (\bibinfo {year}
  {2000})}\BibitemShut {NoStop}%
\bibitem [{\citenamefont {{Zel'dovich}}\ and\ \citenamefont
  {{Polnarev}}(1974)}]{Zeldovich1974}%
  \BibitemOpen
  \bibfield  {author} {\bibinfo {author} {\bibfnamefont {Y.~B.}\ \bibnamefont
  {{Zel'dovich}}}\ and\ \bibinfo {author} {\bibfnamefont {A.~G.}\ \bibnamefont
  {{Polnarev}}},\ }\href@noop {} {\bibfield  {journal} {\bibinfo  {journal}
  {Sov. Ast.}\ }\textbf {\bibinfo {volume} {18}},\ \bibinfo {pages} {17}
  (\bibinfo {year} {1974})}\BibitemShut {NoStop}%
\bibitem [{\citenamefont {Christodoulou}(1991)}]{Christodoulou1991}%
  \BibitemOpen
  \bibfield  {author} {\bibinfo {author} {\bibfnamefont {D.}~\bibnamefont
  {Christodoulou}},\ }\href {\doibase 10.1103/PhysRevLett.67.1486} {\bibfield
  {journal} {\bibinfo  {journal} {Phys. Rev. Lett.}\ }\textbf {\bibinfo
  {volume} {67}},\ \bibinfo {pages} {1486} (\bibinfo {year}
  {1991})}\BibitemShut {NoStop}%
\bibitem [{\citenamefont {Bieri}\ and\ \citenamefont
  {Garfinkle}(2014)}]{BieriGarfinkle2014}%
  \BibitemOpen
  \bibfield  {author} {\bibinfo {author} {\bibfnamefont {L.}~\bibnamefont
  {Bieri}}\ and\ \bibinfo {author} {\bibfnamefont {D.}~\bibnamefont
  {Garfinkle}},\ }\href {\doibase 10.1103/PhysRevD.89.084039} {\bibfield
  {journal} {\bibinfo  {journal} {Phys. Rev. D}\ }\textbf {\bibinfo {volume}
  {89}},\ \bibinfo {pages} {084039} (\bibinfo {year} {2014})}\BibitemShut
  {NoStop}%
\bibitem [{\citenamefont {Strominger}\ and\ \citenamefont
  {Zhiboedov}(2014)}]{Strominger2014}%
  \BibitemOpen
  \bibfield  {author} {\bibinfo {author} {\bibfnamefont {A.}~\bibnamefont
  {Strominger}}\ and\ \bibinfo {author} {\bibfnamefont {A.}~\bibnamefont
  {Zhiboedov}},\ }\href@noop {} {\  (\bibinfo {year} {2014})},\ \Eprint
  {http://arxiv.org/abs/1411.5745} {arXiv:1411.5745 [hep-th]} \BibitemShut
  {NoStop}%
\bibitem [{\citenamefont {{Misner}}\ \emph {et~al.}(1973)\citenamefont
  {{Misner}}, \citenamefont {{Thorne}},\ and\ \citenamefont
  {{Wheeler}}}]{Misner1973}%
  \BibitemOpen
  \bibfield  {author} {\bibinfo {author} {\bibfnamefont {C.~W.}\ \bibnamefont
  {{Misner}}}, \bibinfo {author} {\bibfnamefont {K.~S.}\ \bibnamefont
  {{Thorne}}}, \ and\ \bibinfo {author} {\bibfnamefont {J.~A.}\ \bibnamefont
  {{Wheeler}}},\ }\href@noop {} {\emph {\bibinfo {title} {{Gravitation}}}}\
  (\bibinfo  {publisher} {W. H. Freeman and Co.},\ \bibinfo {address} {San
  Francisco},\ \bibinfo {year} {1973})\BibitemShut {NoStop}%
\bibitem [{\citenamefont {Szabados}(2009)}]{Szabados2009}%
  \BibitemOpen
  \bibfield  {author} {\bibinfo {author} {\bibfnamefont {L.~B.}\ \bibnamefont
  {Szabados}},\ }\href {\doibase 10.12942/lrr-2009-4} {\bibfield  {journal}
  {\bibinfo  {journal} {Living Rev. Relativity}\ }\textbf {\bibinfo {volume}
  {12}},\ \bibinfo {pages} {4} (\bibinfo {year} {2009})}\BibitemShut {NoStop}%
\bibitem [{\citenamefont {Everitt}\ \emph {et~al.}(2011)\citenamefont
  {Everitt}, \citenamefont {DeBra}, \citenamefont {Parkinson}, \citenamefont
  {Turneaure}, \citenamefont {Conklin}, \citenamefont {Heifetz}, \citenamefont
  {Keiser}, \citenamefont {Silbergleit}, \citenamefont {Holmes}, \citenamefont
  {Kolodziejczak}, \citenamefont {Al-Meshari}, \citenamefont {Mester},
  \citenamefont {Muhlfelder}, \citenamefont {Solomonik}, \citenamefont {Stahl},
  \citenamefont {Worden}, \citenamefont {Bencze}, \citenamefont {Buchman},
  \citenamefont {Clarke}, \citenamefont {Al-Jadaan}, \citenamefont
  {Al-Jibreen}, \citenamefont {Li}, \citenamefont {Lipa}, \citenamefont
  {Lockhart}, \citenamefont {Al-Suwaidan}, \citenamefont {Taber},\ and\
  \citenamefont {Wang}}]{PhysRevLett.106.221101}%
  \BibitemOpen
  \bibfield  {author} {\bibinfo {author} {\bibfnamefont {C.~W.~F.}\
  \bibnamefont {Everitt}}, \bibinfo {author} {\bibfnamefont {D.~B.}\
  \bibnamefont {DeBra}}, \bibinfo {author} {\bibfnamefont {B.~W.}\ \bibnamefont
  {Parkinson}}, \bibinfo {author} {\bibfnamefont {J.~P.}\ \bibnamefont
  {Turneaure}}, \bibinfo {author} {\bibfnamefont {J.~W.}\ \bibnamefont
  {Conklin}}, \bibinfo {author} {\bibfnamefont {M.~I.}\ \bibnamefont
  {Heifetz}}, \bibinfo {author} {\bibfnamefont {G.~M.}\ \bibnamefont {Keiser}},
  \bibinfo {author} {\bibfnamefont {A.~S.}\ \bibnamefont {Silbergleit}},
  \bibinfo {author} {\bibfnamefont {T.}~\bibnamefont {Holmes}}, \bibinfo
  {author} {\bibfnamefont {J.}~\bibnamefont {Kolodziejczak}}, \bibinfo {author}
  {\bibfnamefont {M.}~\bibnamefont {Al-Meshari}}, \bibinfo {author}
  {\bibfnamefont {J.~C.}\ \bibnamefont {Mester}}, \bibinfo {author}
  {\bibfnamefont {B.}~\bibnamefont {Muhlfelder}}, \bibinfo {author}
  {\bibfnamefont {V.~G.}\ \bibnamefont {Solomonik}}, \bibinfo {author}
  {\bibfnamefont {K.}~\bibnamefont {Stahl}}, \bibinfo {author} {\bibfnamefont
  {P.~W.}\ \bibnamefont {Worden}}, \bibinfo {author} {\bibfnamefont
  {W.}~\bibnamefont {Bencze}}, \bibinfo {author} {\bibfnamefont
  {S.}~\bibnamefont {Buchman}}, \bibinfo {author} {\bibfnamefont
  {B.}~\bibnamefont {Clarke}}, \bibinfo {author} {\bibfnamefont
  {A.}~\bibnamefont {Al-Jadaan}}, \bibinfo {author} {\bibfnamefont
  {H.}~\bibnamefont {Al-Jibreen}}, \bibinfo {author} {\bibfnamefont
  {J.}~\bibnamefont {Li}}, \bibinfo {author} {\bibfnamefont {J.~A.}\
  \bibnamefont {Lipa}}, \bibinfo {author} {\bibfnamefont {J.~M.}\ \bibnamefont
  {Lockhart}}, \bibinfo {author} {\bibfnamefont {B.}~\bibnamefont
  {Al-Suwaidan}}, \bibinfo {author} {\bibfnamefont {M.}~\bibnamefont {Taber}},
  \ and\ \bibinfo {author} {\bibfnamefont {S.}~\bibnamefont {Wang}},\ }\href
  {\doibase 10.1103/PhysRevLett.106.221101} {\bibfield  {journal} {\bibinfo
  {journal} {Phys. Rev. Lett.}\ }\textbf {\bibinfo {volume} {106}},\ \bibinfo
  {pages} {221101} (\bibinfo {year} {2011})}\BibitemShut {NoStop}%
\bibitem [{\citenamefont {Nichols}\ \emph {et~al.}(2011)\citenamefont
  {Nichols}, \citenamefont {Owen}, \citenamefont {Zhang}, \citenamefont
  {Zimmerman}, \citenamefont {Brink}, \citenamefont {Chen}, \citenamefont
  {Kaplan}, \citenamefont {Lovelace}, \citenamefont {Matthews}, \citenamefont
  {Scheel},\ and\ \citenamefont {Thorne}}]{Nichols2011}%
  \BibitemOpen
  \bibfield  {author} {\bibinfo {author} {\bibfnamefont {D.~A.}\ \bibnamefont
  {Nichols}}, \bibinfo {author} {\bibfnamefont {R.}~\bibnamefont {Owen}},
  \bibinfo {author} {\bibfnamefont {F.}~\bibnamefont {Zhang}}, \bibinfo
  {author} {\bibfnamefont {A.}~\bibnamefont {Zimmerman}}, \bibinfo {author}
  {\bibfnamefont {J.}~\bibnamefont {Brink}}, \bibinfo {author} {\bibfnamefont
  {Y.}~\bibnamefont {Chen}}, \bibinfo {author} {\bibfnamefont {J.~D.}\
  \bibnamefont {Kaplan}}, \bibinfo {author} {\bibfnamefont {G.}~\bibnamefont
  {Lovelace}}, \bibinfo {author} {\bibfnamefont {K.~D.}\ \bibnamefont
  {Matthews}}, \bibinfo {author} {\bibfnamefont {M.~A.}\ \bibnamefont
  {Scheel}}, \ and\ \bibinfo {author} {\bibfnamefont {K.~S.}\ \bibnamefont
  {Thorne}},\ }\href {\doibase 10.1103/PhysRevD.84.124014} {\bibfield
  {journal} {\bibinfo  {journal} {Phys. Rev. D}\ }\textbf {\bibinfo {volume}
  {84}},\ \bibinfo {pages} {124014} (\bibinfo {year} {2011})}\BibitemShut
  {NoStop}%
\bibitem [{\citenamefont {Thorne}(1980)}]{Thorne1980}%
  \BibitemOpen
  \bibfield  {author} {\bibinfo {author} {\bibfnamefont {K.~S.}\ \bibnamefont
  {Thorne}},\ }\href {\doibase 10.1103/RevModPhys.52.299} {\bibfield  {journal}
  {\bibinfo  {journal} {Rev. Mod. Phys.}\ }\textbf {\bibinfo {volume} {52}},\
  \bibinfo {pages} {299} (\bibinfo {year} {1980})}\BibitemShut {NoStop}%
\bibitem [{\citenamefont {Blanchet}(2014)}]{Blanchet2014}%
  \BibitemOpen
  \bibfield  {author} {\bibinfo {author} {\bibfnamefont {L.}~\bibnamefont
  {Blanchet}},\ }\href {\doibase 10.12942/lrr-2014-2} {\bibfield  {journal}
  {\bibinfo  {journal} {Living Rev. Relativity}\ }\textbf {\bibinfo {volume}
  {17}},\ \bibinfo {pages} {2} (\bibinfo {year} {2014})}\BibitemShut {NoStop}%
\bibitem [{\citenamefont {Vines}\ and\ \citenamefont
  {Nichols}(2014)}]{Vines2014}%
  \BibitemOpen
  \bibfield  {author} {\bibinfo {author} {\bibfnamefont {J.}~\bibnamefont
  {Vines}}\ and\ \bibinfo {author} {\bibfnamefont {D.~A.}\ \bibnamefont
  {Nichols}},\ }\href@noop {} {\  (\bibinfo {year} {2014})},\ \Eprint
  {http://arxiv.org/abs/1412.4077} {arXiv:1412.4077 [gr-qc]} \BibitemShut
  {NoStop}%
\bibitem [{\citenamefont {Flanagan}\ \emph {et~al.}(2015)\citenamefont
  {Flanagan}, \citenamefont {Harte},\ and\ \citenamefont
  {Nichols}}]{Flanagan2015}%
  \BibitemOpen
  \bibfield  {author} {\bibinfo {author} {\bibfnamefont {{\'E}.~{\'E}.}\
  \bibnamefont {Flanagan}}, \bibinfo {author} {\bibfnamefont {A.~I.}\
  \bibnamefont {Harte}}, \ and\ \bibinfo {author} {\bibfnamefont {D.~A.}\
  \bibnamefont {Nichols}},\ }\href@noop {} {\enquote {\bibinfo {title}
  {unpublished},}\ } (\bibinfo {year} {2015})\BibitemShut {NoStop}%
\bibitem [{\citenamefont {Pasterski}\ \emph {et~al.}(2015)\citenamefont
  {Pasterski}, \citenamefont {Strominger},\ and\ \citenamefont
  {Zhiboedov}}]{Pasterski2015}%
  \BibitemOpen
  \bibfield  {author} {\bibinfo {author} {\bibfnamefont {S.}~\bibnamefont
  {Pasterski}}, \bibinfo {author} {\bibfnamefont {A.}~\bibnamefont
  {Strominger}}, \ and\ \bibinfo {author} {\bibfnamefont {A.}~\bibnamefont
  {Zhiboedov}},\ }\href@noop {} {\  (\bibinfo {year} {2015})},\ \Eprint
  {http://arxiv.org/abs/1502.06120} {arXiv:1502.06120 [hep-th]} \BibitemShut
  {NoStop}%
\bibitem [{\citenamefont {{Winicour}}(2014)}]{Winicour2014}%
  \BibitemOpen
  \bibfield  {author} {\bibinfo {author} {\bibfnamefont {J.}~\bibnamefont
  {{Winicour}}},\ }\href {\doibase 10.1088/0264-9381/31/20/205003} {\bibfield
  {journal} {\bibinfo  {journal} {Classical Quantum Gravity}\ }\textbf
  {\bibinfo {volume} {31}},\ \bibinfo {eid} {205003} (\bibinfo {year}
  {2014})}\BibitemShut {NoStop}%
\bibitem [{\citenamefont {Vines}(2014)}]{Vines2014b}%
  \BibitemOpen
  \bibfield  {author} {\bibinfo {author} {\bibfnamefont {J.}~\bibnamefont
  {Vines}},\ }\href@noop {} {\enquote {\bibinfo {title} {private
  communication},}\ } (\bibinfo {year} {2014})\BibitemShut {NoStop}%
\end{thebibliography}%
\end{document}